\documentclass[aps,showpacs,twocolumn,
superscriptaddress]{revtex4}
\usepackage{graphicx}
\usepackage{dcolumn}
\usepackage{bm}
\usepackage{color}
\usepackage[normalem]{ulem} 
\usepackage[dvipsnames]{xcolor} 
\usepackage{hyperref}
\hypersetup{
  colorlinks=true,        
  linkcolor=cyan,         
  citecolor=cyan,         
}
\usepackage{mathrsfs}
\usepackage[utf8]{inputenc}
\usepackage{mathtools}
\usepackage{doi}
\usepackage{amsmath}
\usepackage{amssymb}

\usepackage{soul}

\usepackage{scalerel}
\usepackage{tikz}
\usetikzlibrary{svg.path}

\definecolor{orcidlogocol}{HTML}{A6CE39}
\tikzset{
  orcidlogo/.pic={
    \fill[orcidlogocol] svg{M256,128c0,70.7-57.3,128-128,128C57.3,256,0,198.7,0,128C0,57.3,57.3,0,128,0C198.7,0,256,57.3,256,128z};
    \fill[white] svg{M86.3,186.2H70.9V79.1h15.4v48.4V186.2z}
                 svg{M108.9,79.1h41.6c39.6,0,57,28.3,57,53.6c0,27.5-21.5,53.6-56.8,53.6h-41.8V79.1z M124.3,172.4h24.5c34.9,0,42.9-26.5,42.9-39.7c0-21.5-13.7-39.7-43.7-39.7h-23.7V172.4z}
                 svg{M88.7,56.8c0,5.5-4.5,10.1-10.1,10.1c-5.6,0-10.1-4.6-10.1-10.1c0-5.6,4.5-10.1,10.1-10.1C84.2,46.7,88.7,51.3,88.7,56.8z};
  }
}

\newcommand\orcidicon[1]{\href{https://orcid.org/#1}{\mbox{\scalerel*{
\begin{tikzpicture}[yscale=-1,transform shape]
\pic{orcidlogo};
\end{tikzpicture}
}{|}}}}

\usepackage{orcidlink}    

\begin{document}

\title{Chaotic motion and power spectral density in Schwarzschild Bertotti-Robinson black hole spacetime}

\author{Yunqiao Xu}
\email{25B911033@stu.hit.edu.cn}
\affiliation{School of Physics, Harbin Institute of Technology, Harbin 150001, People’s Republic of China}

\author{Uktamjon Uktamov\orcidlink{0009-0003-0423-2474}} 
\email[Corresponding Author:]{uktam.uktamov11@gmail.com}
\affiliation{School of Physics, Harbin Institute of Technology, Harbin 150001, People’s Republic of China}
\affiliation{University of Tashkent for Applied Sciences, Str. Gavhar 1, Tashkent 100149, Uzbekistan}
\affiliation{Institute for Advanced Studies, New Uzbekistan University, Movarounnahr str. 1, Tashkent 100000, Uzbekistan}

\author{Pierros Ntelis \orcidlink{0000-0002-7849-2418}}
	\email{ntelis.pierros@gmail.com}
    \affiliation{School of Physics, Harbin Institute of Technology, Harbin 150001, People’s Republic of China}
	\affiliation{Institute of Theoretical Physics, National University of Uzbekistan, Tashkent 100174, Uzbekistan}

\author{Ahmadjon~Abdujabbarov,\orcidlink{0000-0002-6686-3787}}
	\email{ahmadjon@astrin.uz}

    \affiliation{School of Physics, Harbin Institute of Technology, Harbin 150001, People’s Republic of China}
    
    \affiliation{Institute of Fundamental and Applied Research, National Research University TIIAME, Kori Niyoziy 39, Tashkent 100000, Uzbekistan}
    
    \affiliation{Tashkent State Technical University, Tashkent 100095, Uzbekistan}

	\author{Bobomurat Ahmedov\orcidlink{0000-0002-1232-610X}}
	\email{ahmedov@astrin.uz}
    \affiliation{School of Physics, Harbin Institute of Technology, Harbin 150001, People’s Republic of China}
	\affiliation{Institute of Theoretical Physics, National University of Uzbekistan, Tashkent 100174, Uzbekistan}

 \author{Chengxun Yuan\orcidlink{0000-0002-2308-6703}}
    \email{yuancx@hit.edu.cn}

    \affiliation{School of Physics, Harbin Institute of Technology, Harbin 150001, People’s Republic of China}

\date{\today}

\begin{abstract}
In this paper, we show that in weak field limit Schwarzschild Bertotti-Robinson black hole (Schwarzschild-BR BH) turns into Schwarzschild black hole immersed in external uniform magnetic field which is given in \cite{Wald:1984rg}. The dynamics of both magnetized and electrically charged particles in the vicinity of a Schwarzschild-BR black hole are investigated. The innermost stable circular orbits (ISCOs) for both magnetized and electrically charged particles are examined in detail, revealing that the magnetic field parameter $B$ exerts a considerable influence, leading to an increase in the ISCO radius. The orbital and epicyclic motion of test particles in Schwarzschild-BR black hole spacetime was analyzed, including both circular orbits and their oscillatory perturbations. Additionally, the trajectories of both magnetized and electrically charged particles are analyzed for various configurations of the magnetic parameter $B$. We also demonstrate how the magnetic field $B$, electric charge $q$, and magnetic moment $\mu$ influence the dynamics of charged particles, specifically affecting the chaotic behavior, Poincaré sections, oscillatory frequencies and power spectral density.
\end{abstract}

\maketitle

\tableofcontents

\section{Introduction}

The interplay between strong gravitational fields and electromagnetic phenomena is a cornerstone of modern relativistic astrophysics. Black holes, the most compact objects predicted by General Relativity (GR), are often not isolated but embedded in complex environments, including plasma, accretion disks, and large-scale magnetic fields \cite{Wald:1984rg}. Understanding how these external fields modify the spacetime geometry and influence the motion of test particles is crucial for interpreting high-energy observations, from quasi-periodic oscillations (QPOs) in X-ray binaries to the dynamics of stars near the Galactic center \cite{Uktamov:2025bth, Xamidov:2025hrj}. While GR provides the foundational framework, the extreme conditions near black holes also serve as a unique laboratory to probe potential deviations from Einstein's theory and to explore more fundamental descriptions of gravity and their coupling to matter.

A particularly elegant method to study these interactions is through exact solutions of the Einstein-Maxwell equations. The Schwarzschild metric, describing a static, uncharged black hole, is a fundamental solution. However, embedding it in a background magnetic field breaks this simplicity. While the Schwarzschild-Melvin solution \cite{Cardoso:2024yrb} describes a black hole in a unidirectional magnetic field, a compelling alternative is provided by the Schwarzschild black hole immersed in a uniform Bertotti-Robinson (BR) magnetic field, recently derived in \cite{Podolsky:2025tle}. This new exact solution, which we analyze in this work, offers a rich, non-asymptotically flat spacetime where the magnetic field is not merely a test field but actively participates in shaping the curvature, leading to a confining, universe-like topology in the external region. This provides a theoretically clean and tractable model to explore magnetized environments beyond the test-field approximation.

The study of particle dynamics in such backgrounds is the primary tool for extracting observable signatures. The motion of both magnetized particles (possessing a magnetic dipole moment) and electrically charged particles is governed by the interplay of gravitational and electromagnetic forces. For magnetized particles, the interaction with the gradient of the magnetic field leads to an effective potential that can significantly alter orbital stability and produce epicyclic resonances \cite{2003CQGra..20..469D, Uktamov:2024zmj}. For charged particles, the Lorentz force couples directly to the four-potential of the BR field, leading to complex trajectories that can range from regular to chaotic \cite{Kolos:2023oii}. The motion of particles with zero or nonzero electric and magnetic charges around compact objects within various gravitational frameworks has been investigated in Refs.~\cite{AaJuraeva2021,AaRayimbaev2023,AaRahimov2011499, AaNarzilloev2021,AaVrba2020,AaRayimbaev2020,AaAbdujabbarov2013,AaShaymatov2021,AaRayimbaev2021,AaHaydarov2020}.
The energy characteristics near black holes using diverse approaches have been analyzed in~\cite{AaAbdujabbarov2011,AaTursunov2013,AaShaymatov2018,AaAbdujabbarov2013173,AaShaymatov2013,AaToshmatov2015}.
Gravitational lensing and photon trajectories around black holes in modified gravity theories have been examined in Refs.~\cite{AaBenavides-Gallego2018,AaDitta2023,AaTurimov2019,AaSarikulov2022}. A detailed analysis of these dynamics, including the identification of stable circular orbits, the innermost stable circular orbit (ISCO), and the characteristic epicyclic frequencies, is essential for building accurate models of accretion disks and QPOs.

The observational signatures of black holes are profoundly shaped by their surrounding environments. While our work focuses on the influence of a uniform magnetic field on particle dynamics, a parallel and highly active line of inquiry concerns the effects of dark matter and plasma. 

The study of black holes in realistic astrophysical environments has witnessed remarkable progress in recent years, with numerous investigations exploring how various physical fields and surrounding matter affect their observational signatures. A substantial body of work has been devoted to understanding black hole horizons, acceleration radiation, and nonlinear electrodynamic effects \cite{Uktamov:2602.15077, Uktamov:2512.21387,Yang:2603.15241}, as well as the detailed dynamics of charged and spinning particles around magnetized black holes, with direct applications to S2 star observations and hotspot phenomena in the Galactic center \cite{Uktamov:2510.16315, Uktamov:2025bth, Uktamov:2024zmj}. The influence of dark matter halos on black hole spacetimes has emerged as a particularly active research frontier, with studies examining static and rotating black holes embedded in various dark matter profiles, their constraints from Event Horizon Telescope observations, and their imprint on quasiperiodic oscillations and accretion disk properties \cite{Uktamov:2509.00460, Uktamov:2505.20031, Xamidov:2025hrj, Alloqulov:2504.01651, Alloqulov:EPJC85-798}. Complementary to these efforts, the role of plasma environments has been systematically investigated through gravitational lensing studies, neutrino oscillations, and the optical properties of black holes in both uniform and non-uniform plasma distributions \cite{Alloqulov:2512.12672, Khasanov:JCAP2025, Alloqulov:EPJC86-208, Mannobova:PDU2025, Kholmuminov:NuPhB2025}. The gravitational wave signatures of black holes in modified gravity theories and exotic backgrounds have also been extensively explored, particularly through periodic orbits and extreme mass ratio inspirals \cite{Alloqulov:EPJC86-259, Alloqulov:EPJC86-117, Mirkhaydarov:ChinPhysC2025, Alloqulov:2512.12672}. On the theoretical front, foundational frameworks such as Functors of Actions theories and analytical poly$\Lambda$CDM dynamics have been developed to systematically classify gravitational actions and understand cosmological evolution through dynamical systems techniques \cite{ Ntelis2023, Ntelis2024, Ntelis2025a_ATT, ntelis2025b, Ntelis:IJGMMP2025, Ntelis:IJGMMP2025b,Bahamonde:PDU2025,DiValentino:NatureAstron2025,Capozziello:PhysRep2011}. These theoretical advances are complemented by ongoing efforts to address observational tensions in cosmology through systematic studies of large-scale structure and fundamental physics \cite{DiValentino:PDU2025, Ntelis:JCAP2018, Ntelis:JCAP2017}. Within this rich landscape, our work on the Schwarzschild-Bertotti-Robinson black hole occupies a unique niche: while previous studies have examined black holes in magnetic fields \cite{Uktamov:2024ckf, Uktamov:2024zmj}, dark matter halos \cite{Uktamov:2509.00460, Alloqulov:EPJC85-798}, and plasma environments \cite{Alloqulov:2512.12672, Khasanov:JCAP2025} separately, our investigation provides a comprehensive analysis of both magnetized and charged particle dynamics in an exact solution where the magnetic field is not a test field but actively shapes the spacetime curvature. By systematically studying the ISCO parameters, epicyclic frequencies, and trajectory stability—including chaos regularization through Poincaré sections and power spectral density analysis—we bridge the gap between abstract theoretical frameworks and concrete observational predictions, demonstrating how the SBR spacetime serves as a valuable laboratory for understanding black hole environments beyond the test-field approximation.

This work sits at the intersection of concrete phenomenological modeling and advanced theoretical frameworks. The system under investigation—a Schwarzschild black hole immersed in a Bertotti-Robinson magnetic field—is an exact solution to the standard Einstein-Maxwell equations. As such, it serves as a perfect physical realization of the principles underlying more abstract formalisms like Functors of Actions theories (FAT) \cite{Ntelis2023, Ntelis2024}. The FAT framework provides a categorical language to construct and classify gravitational actions; our specific model corresponds to the limit where the general FAT action simplifies to its standard, well-known form. In this sense, our work provides a concrete example of the physical spacetimes that emerge from the FAT paradigm when one restricts to classical general relativity. Furthermore, our detailed analysis of particle dynamics relies on standard tensor calculus, which forms the essential foundation for more complex advanced tensor theories (ATT) \cite{Ntelis2025a_ATT}. While ATT extends this calculus to incorporate elements like fractional operators, the core geometrical and field-theoretic objects we manipulate (metrics, connection, curvature, and Maxwell tensors) are the basic building blocks of that framework. Consequently, this study not only offers new astrophysical insights into magnetized black hole environments but also establishes a well-understood physical baseline for future investigations that could generalize the action or the tensor calculus within the richer landscapes of FAT and ATT.

Beyond these foundational connections, the analytical methods employed in this study resonate with techniques developed in cosmological contexts. The detailed phase space analysis of particle orbits—identifying fixed points (ISCOs), studying their stability, and mapping regions of regular versus chaotic behavior—finds a striking methodological parallel in analytical poly$\Lambda$CDM dynamics \cite{ntelis2025b}. Just as that framework explores the dynamical systems governing cosmological evolution, identifying attractors and analyzing stability in the expansion history of the universe, we here perform an analogous analysis for test particles in a strong-field gravitational environment. This parallel underscores the universality of dynamical systems techniques across different scales in gravitational physics, from the cosmic horizon to the black hole event horizon.

In this paper, we present a comprehensive study of particle dynamics in the Schwarzschild-BR black hole spacetime. We begin by analyzing the spacetime structure and the topology of the magnetic field lines as measured by zero-angular-momentum observers (ZAMOs). Subsequently, we investigate two distinct classes of particles:

\begin{enumerate}
    \item Magnetized particles, deriving the equations of motion, effective potential, and conditions for circular orbits. We compute the ISCO parameters as functions of the magnetic field $B$ and the magnetic coupling parameter $\beta$, and analyze the trajectory's regularity via Poincar\'e sections and power spectral densities. The epicyclic frequencies ($\nu_r, \nu_\theta, \nu_K$) are derived and their radial profiles are explored.
    
    \item Electrically charged particles, for which we derive the Hamilton-Jacobi equation and the effective potential. We perform a similar analysis of circular orbits, ISCO parameters, and trajectory stability, highlighting the distinct roles of the charge-to-mass ratio $\beta_E$ and the magnetic field $B$. The epicyclic frequencies for charged particles are also derived using a perturbative approach.
\end{enumerate}

Our results reveal that both the magnetic field $B$ and the particle's coupling parameters ($\beta$, $\beta_E$) have a profound impact on the orbital dynamics, generally acting to regularize chaotic motion and alter the characteristic frequency spectrum. This work not only provides new insights into the Schwarzschild-BR spacetime but also demonstrates how exact solutions can serve as a bridge between phenomenological astrophysics and fundamental theoretical frameworks.

The paper is organized as follows. In Section \ref{sec:SBRBH}, we introduce the SBR black hole spacetime, presenting the metric, the four-potential, and the electromagnetic field tensor. We analyze the spacetime structure, derive the magnetic field components as measured by ZAMOs, and visualize the magnetic field lines for different field strengths. We also examine the horizon structure and the asymptotic behavior of the magnetic field, recovering the Wald solution in the weak-field limit. Section \ref{sec:Motion_of_particles_with_Magnetic_dipole_momentum} is devoted to the dynamics of particles possessing a magnetic dipole moment. We derive the equations of motion, the effective potential, and the conditions for circular orbits. The ISCO parameters are computed as functions of the magnetic field $B$ and the magnetic coupling parameter $\beta$. Subsequently, in Section \ref{sec:Epicyclic_Frequencies_for_Bparticles_in_SBR_BH}, we calculate the epicyclic frequencies for magnetized particles and analyze their radial profiles. Within this subsection, Section \ref{sec:trajectory_dynamics_interpretation} provides a detailed interpretation of the trajectory dynamics, including power spectral density analysis and Poincaré sections to characterize regular and chaotic motion, while Section \ref{sec:Harmonic_oscillation_frequencies} presents the harmonic oscillation frequencies and their dependence on the spacetime parameters. In Section IV, we investigate the motion of electrically charged particles around the SBR black hole. We derive the Hamilton-Jacobi equation, the effective potential, and the ISCO parameters, highlighting the role of the charge-to-mass ratio $\beta_E$. Section \ref{sec:Motion_of_electrically_charged_particles} extends this analysis to the epicyclic frequencies of charged particles. Finally, Section \ref{sec:conclusion} summarizes our main findings and conclusions, while Section \ref{sec.app} contains the Appendix with supplementary mathematical expressions, followed by the References.

\section{Schwarzschild Bertotti-Robinson black hole}\label{sec:SBRBH}
\begin{figure*}[ht!]
\includegraphics[width=0.4\textwidth]{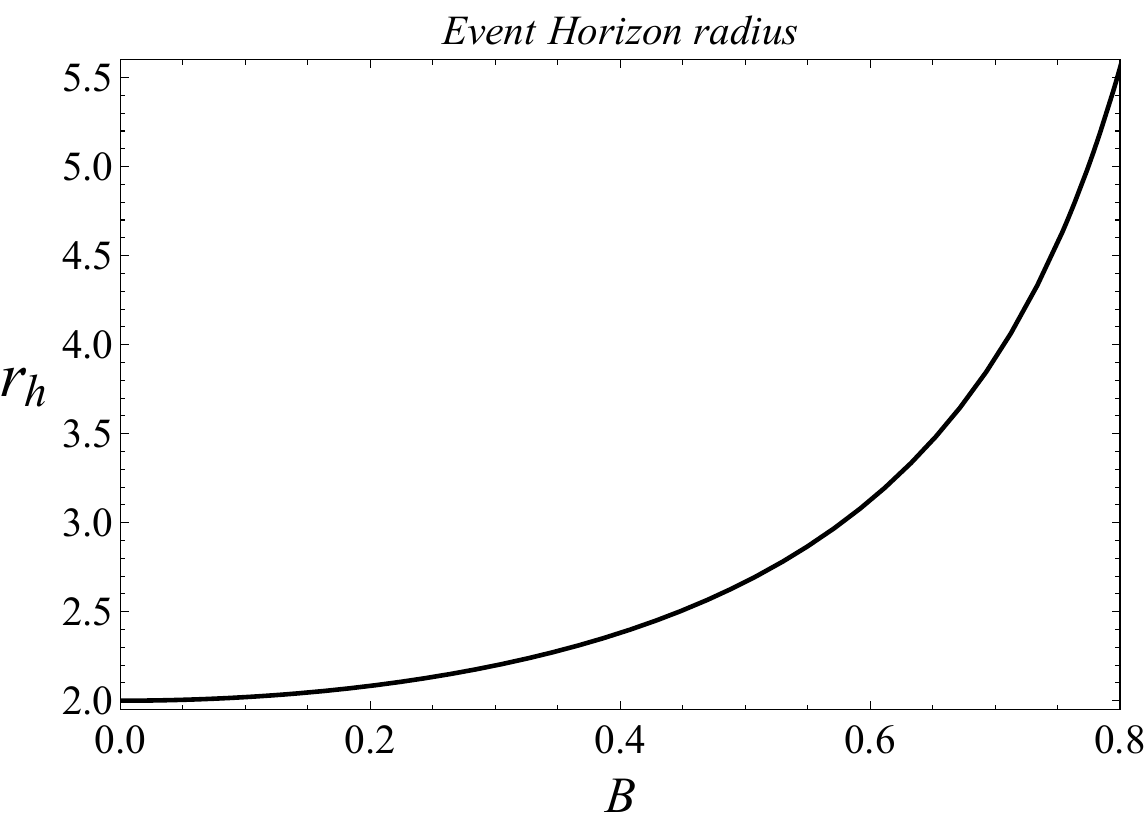}
\caption{The variation of event horizon radius with magnetic field strength B in equatorial orbit.\label{Fig.r_h}}
\end{figure*}

\begin{figure*}[ht!]
\includegraphics[width=0.3\textwidth]{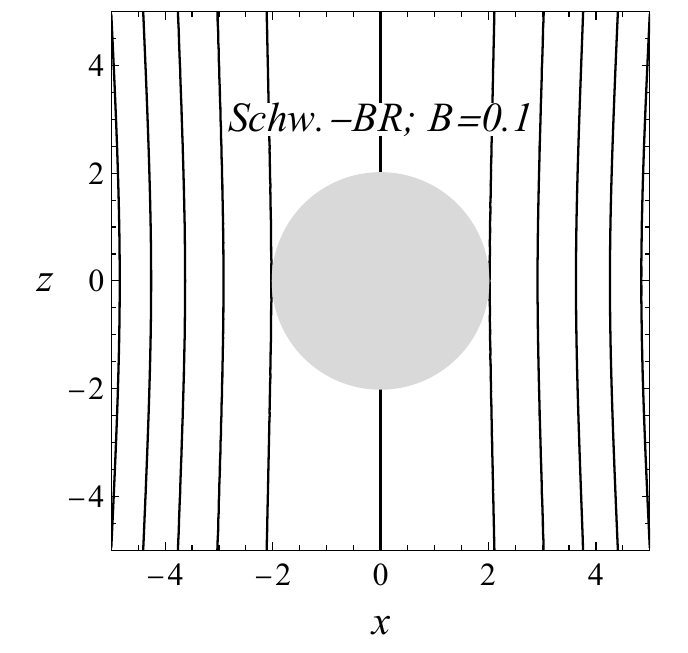}
\includegraphics[width=0.3\textwidth]{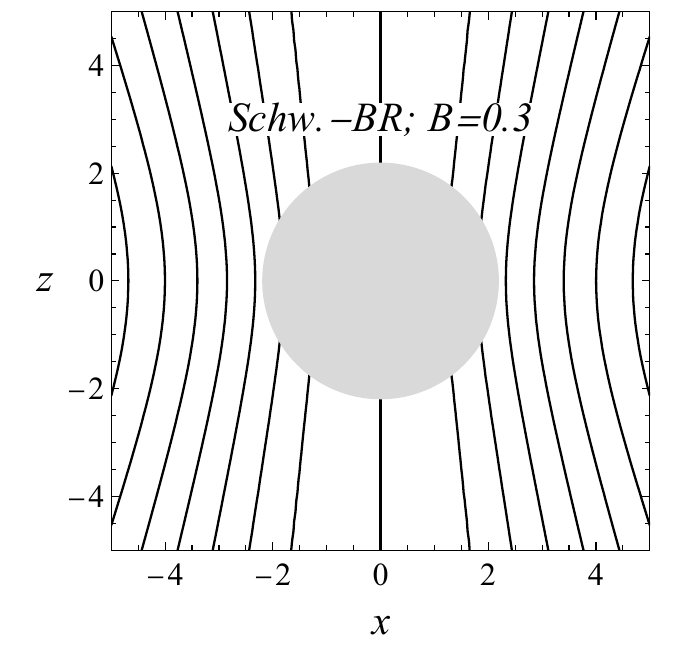}
\includegraphics[width=0.3\textwidth]{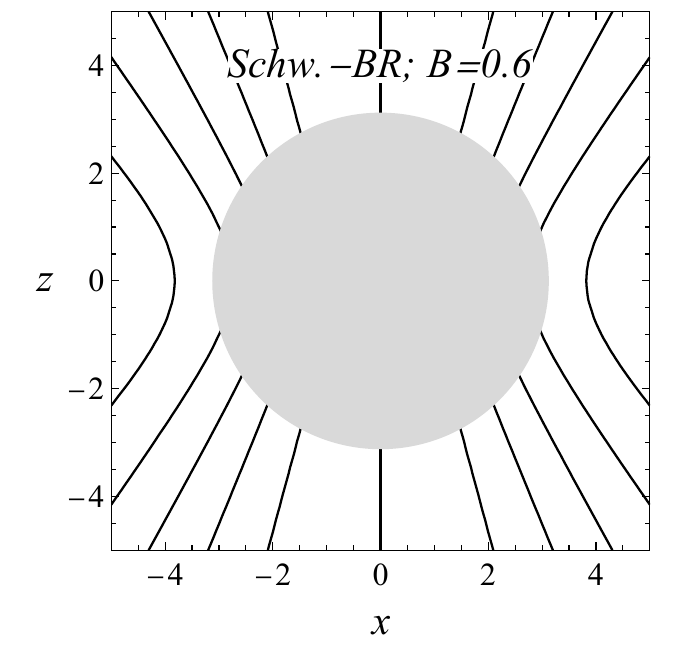}
\caption{The plot shows the magnetic field lines in the vicinity of the  Schwarzschild BR.\label{Fig.B}}
\end{figure*}

We consider a Schwarzschild black hole with mass $M$ immersed in an external asymtotically uniform Bertotti-Robinson magnetic field $B$, with the new line element given by~\cite{Podolsky:2025tle}:
\begin{equation}\label{eq.metric}
  ds^2=\frac{1}{R^2}[-Qdt^2+\frac{dr^2}{Q}+r^2(\frac{d\theta^2}{P}+P\sin^2{\theta}d\phi^2)]  , 
\end{equation}
where
\begin{equation}
P = 1 + B^2 M^2 \cos^2 \theta\,,
\end{equation}
\begin{equation}
Q = \left( 1 + B^2 r^2 \right) \left( 1 - B^2 M^2 - \frac{2M}{r} \right)\,,
\end{equation}
\begin{equation}
R^2 = 1 + B^2 \left[ r^2 \sin^2 \theta + \left( 2M r + B^2 M^2 r^2 \right) \cos^2 \theta \right]\,.
\end{equation}
According to approach given by~\cite{Podolsky:2025tle},there is only one non-vanishing component of the four-potential:
\begin{equation}\label{eq.Aphi}
A_\phi = \frac{1}{B}(\frac{B^2 M r \cos ^2\theta+1}{R}-1).
\end{equation}
Then  non-vanishing components of the electromagnetic field tensor $F_{\mu\nu}=\partial_\mu A_\nu-\partial_\nu A_\mu$ can be calculated in the following form
\begin{equation}\label{eq.Frphi}
F_{r\phi}=-\frac{Br\sin ^2\theta\left[B^2 M^2 \cos (2 \theta)+B^2 M^2+2\right]}{2 R^3}\,,
\end{equation}

\begin{multline}\label{eq.Fthetaphi}
F_{\theta\phi}=\frac{1}{8 R^3}\bigg\{Br^2 \big[B^2 M \sin4 \theta\left(r-2 M -B^2 M^2 r\right)\\
-2 \sin2 \theta \left(B^4 M^3 r+3 B^2 M r+2\right)\big]\bigg\}\,.
\end{multline}
To measure the magnetic field locally, we use the orthonormal tetrad of a ZAMO. The magnetic field components are calculated using the Levi-Civita tensor~\cite{2003CQGra..20..469D, Uktamov:2024ckf}:
\begin{eqnarray}\label{eq.fields}
B^i &=& \frac{1}{2}\eta^{i \beta \sigma \mu} F_{\beta \sigma} w_{\mu},
\end{eqnarray}
where $i=(r,\theta,\phi)$, $\bm{w}$ represents the four-velocity of the ZAMOs with $w^{\mu}={R}/{\sqrt{Q}}(1,0,0,0)$, and the symbol $\eta_{\alpha\beta\sigma\gamma}$
represents the tensor form of the Levi-Civita symbol $\epsilon_{\alpha\beta\sigma\gamma}$, satisfying the following relationships
\begin{eqnarray}
&&\eta_{\alpha \beta \sigma \gamma}=\sqrt{-g}\,\epsilon_{\alpha \beta \sigma \gamma},\quad \eta^{\alpha \beta \sigma \gamma}=-\frac{1}{\sqrt{-g}}\epsilon^{\alpha \beta \sigma \gamma},
\end{eqnarray}
where $g={{\rm{det}}|g_{\mu \nu}|}=-{r^4\sin^2\theta}/{R^8}$ is the determinant the spacetime metric (\ref{eq.metric}). Subsequently, we can find the components of the magnetic field by employing Eqs.~(\ref{eq.Frphi}, \ref{eq.Fthetaphi}, \ref{eq.fields}) as:
\begin{eqnarray}\label{eq.Br}
B^{\hat{r}}&=&\frac{B\cos\theta }{2R}\big\{B^4 M^3r+3 B^2 M r+2\\\nonumber
&+&B^2M\cos 2\theta \left[M \left(B^2 M r+2\right)-r\right]\big\}\,,
\end{eqnarray}
\begin{equation}\label{eq.Btheta}
B^{\hat{\theta}}=-\frac{B\sin\theta}{R}\sqrt{PQ}\,.
\end{equation}
We have shown dependence of the horizon radius $r_h$ on magnetic field $B$ in Fig.~\ref{Fig.r_h} and also  magnetic field lines in the vicinity of the Schwarzschild BR BH is illustrated in Fig.~\ref{Fig.B}. One can conclude from these plots in Fig.~\ref{Fig.B} that increasing the value of the magnetic filed $B$ causes enlarging horizon radius of the Schwarzschild BR BH. 

\subsection{Analyzing space-time structure}
One can find the horizon of the Schwarzschild-BR BH using condition $Q=0$ in the form:
\begin{eqnarray}\label{eq.r_h}
    r_h=\frac{2M}{1-B^2M^2}\,,
\end{eqnarray}
so maximum value of the magnetic field $B={1}/{M}$ of the Schwarzschild-BR BH can be obtained.  

Also, employing Eqs.~(\ref{eq.Br}, \ref{eq.Btheta}) asymptotic behavior ($r\to\infty$) of the magnetic field can be analyzed:
\begin{eqnarray}\label{eq.B asm}
B^{\hat{r}}=\frac{B^2M\cos{\theta}\left[B^2M^2\cos^2{\theta}+\sin^2{\theta}+1\right]}{\sqrt{\sin^2{\theta}+B^2M^2\cos^2{\theta}}}\,,
\end{eqnarray}
\begin{eqnarray}\label{eq.B asm2}
    B^{\hat{\theta}}=-\frac{B\sin{\theta}\sqrt{1-B^2M^2}}{\sqrt{\sin^2{\theta}+B^2M^2\cos^2{\theta}}}\,,
\end{eqnarray}
then if we recalculate Eqs.(\ref{eq.B asm},\ref{eq.B asm2}) for extreme case $B\to{1}/{M}$ we will get:
\begin{eqnarray}\label{eq.extr B}
&B^{\hat{r}}\to2B\cos{\theta}\,,
\end{eqnarray}
\begin{eqnarray}
 B^{\hat{\theta}}\to0\,.   
\end{eqnarray}
Actually, the astrophysical allowed value of the magnetic field is $B\ll1$, for example if we calculate the extreme value of the $B$ for $Sgr A^*$:
\begin{eqnarray}\label{eq.extreme B}
    B=\frac{1}{M^*}\frac{c^4}{G^{{3}/{4}}}\approx5.49024\times10^{12} \text{Gauss}\,,
\end{eqnarray}
where $M^*$ is the mass of the $Sgr A^*$. Thus, Eqs.~(\ref{eq.Br},\ref{eq.Btheta}) can be approximated as:
\begin{eqnarray}\label{eq.B app}
    B^{\hat{r}}=B\cos{\theta}\left(1+\mathcal{O}(B^2)\right)\,,
\end{eqnarray}
\begin{eqnarray}\label{eq.B app2}
    B^{\hat{\theta}}=-B\sin{\theta}\sqrt{1-\frac{2M}{r}}\left(1+\mathcal{O}(B^2)\right)\,,
\end{eqnarray}
which is exactly the same with Wald's approach~(see, e.g. \cite{Wald:1984rg, Uktamov:2024zmj}). Also, we should note that four-potential (\ref{eq.Aphi}) of the Schwarzschild-BR BH can be approximated as:
\begin{eqnarray}\label{eq.Aphi2}
    A_\phi=-\frac{1}{2}Br^2\sin^2{\theta}+\mathcal{O}(B^3)\,,
\end{eqnarray}
which is again four-potential of the Schwarzschild BH immersed in external magnetic field derived by Wald~\cite{Wald:1984rg}.

\section{The motion of the particles with magnetic dipole momentum}\label{sec:Motion_of_particles_with_Magnetic_dipole_momentum}

The Lagrangian for the magnetized particle with the rest mass $m$ in the vicinity Schwarzschild BR BH  can be expressed as~\cite{2003CQGra..20..469D}:

\begin{eqnarray}\label{5}
    \mathscr{L}=\frac{1}{2}(m+U)g_{\mu\nu}u^\mu u^\nu-\frac{1}{2} U,
\end{eqnarray}
where $U=D^{\mu\nu}F_{\mu\nu}$ is the magnetic interaction term.  If we restrict the motion of the magnetized particles to the equatorial plane and consider their magnetic dipole to be aligned orthogonally to it, the four-vector of the magnetic moment simplifies to $\mu^{\hat{\alpha}} = (0,0,\mu,0)$, where $\mu$ is the absolute value of the magnetic momentum of the particle. Subsequently, the conjugate four momenta can be expressed as 
\begin{eqnarray}\label{6}
  p_\mu=\frac{\partial\mathscr{L}}{\partial \Dot{x}^\mu}=(m+U)g_{\mu\nu}u^\nu,  
\end{eqnarray}
so magnetized particle's specific energy $\mathcal{E}=E/m$ and specific angular momentum $l=L/m$ are:
\begin{eqnarray}\label{eq.energy}
    -\mathcal{E}&=&\left(1-\beta\frac{\sin\theta}{R}\sqrt{PQ}\right)g_{tt}\Dot{t}\,,\\ \label{eq.angular momen.}
    l&=&\left(1-\beta\frac{\sin\theta}{R}\sqrt{PQ}\right)g_{\phi\phi}\Dot{\phi},\,,
\end{eqnarray}
here $\beta=\mu B/m$ is the magnetic coupling parameter. Furthermore, Hamilton-Jacobi equation for particles with magnetic dipole momentum can be expressed as:
\begin{eqnarray}\label{8}
   g^{\mu\nu}\frac{\partial {\cal S}}{\partial x^\mu}\frac{\partial {\cal S}}{\partial x^\nu}=-m^2\left(1-\frac{U}{m}\right)^2. 
\end{eqnarray}
 where ${\cal S}=-Et+L\phi+{\cal S}_r(r)$ is the Jacobi action. Consequently, the equation governing radial motion takes the simple form:
\begin{eqnarray}\label{9}
    \dot{r}^2=\frac{R^4}{\left[1-\beta f(r)\right]^2}\left(\mathcal{E}^2-V_{\mathrm{eff}}\right),
\end{eqnarray}
in which
\begin{eqnarray}\label{eq.Veff1}
V_{\mathrm{eff}}(r;l)=Q\left[\frac{l^{2}}{r^{2}}+\frac{1}{1+B^{2}r^{2}}\{1+\beta f(r)\}^{2}\right].
\end{eqnarray}

\begin{figure*}[ht!]
\includegraphics[width=0.4\textwidth]{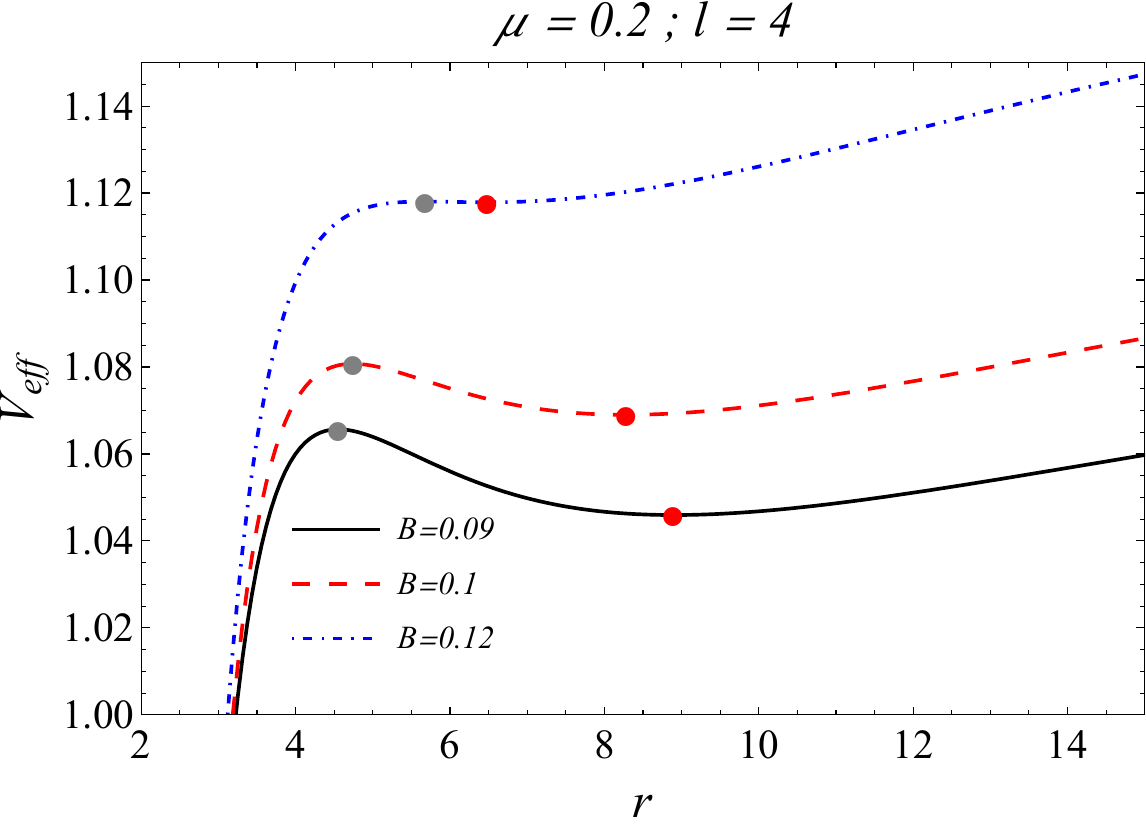}
\includegraphics[width=0.4\textwidth]{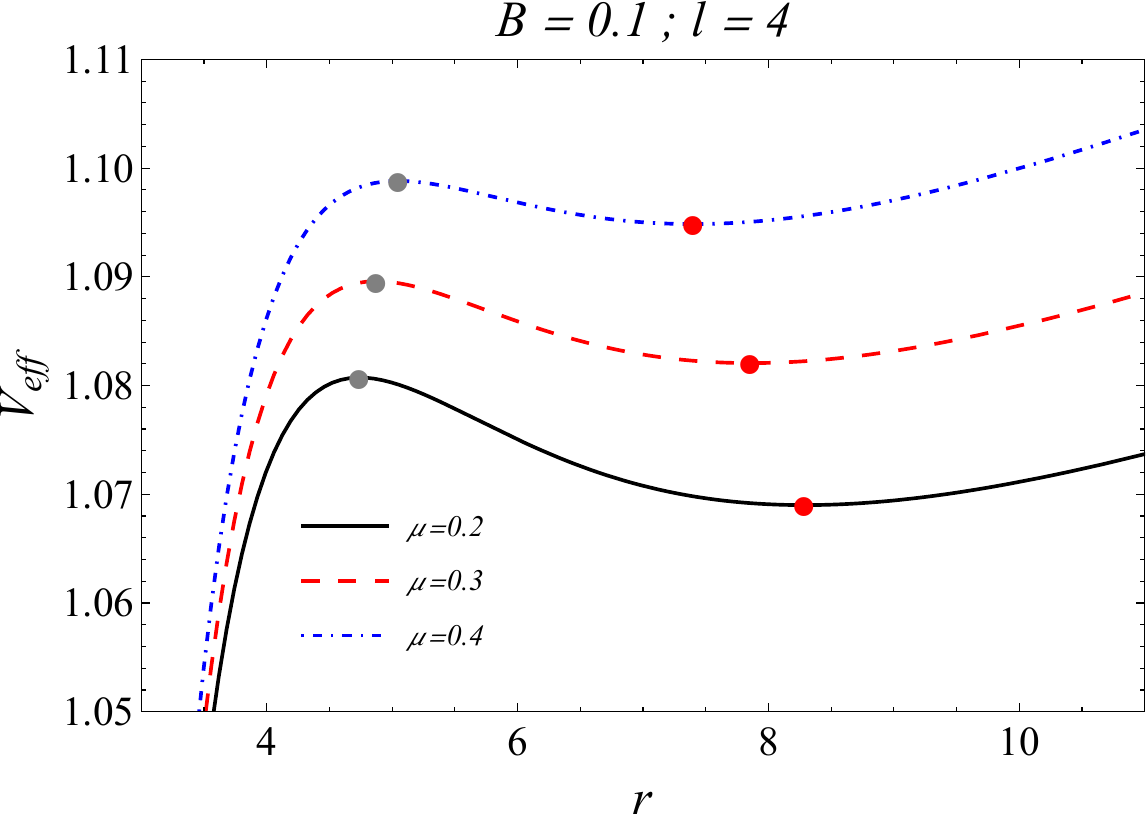}
\caption{The radial profile of the effective potential is depicted for fixed values of the magnetic moment $\mu$ (left panel), and
the magnetic field $B$ (right panel). The minima ($r_{\text{min}}$) in the effective potential correspond to stable circular orbits, while
the maxima ($r_{\text{max}}$) correspond to unstable circular orbits.\label{Fig.effective}}
\end{figure*}

\begin{figure*}[ht!]
\includegraphics[width=0.3\textwidth]{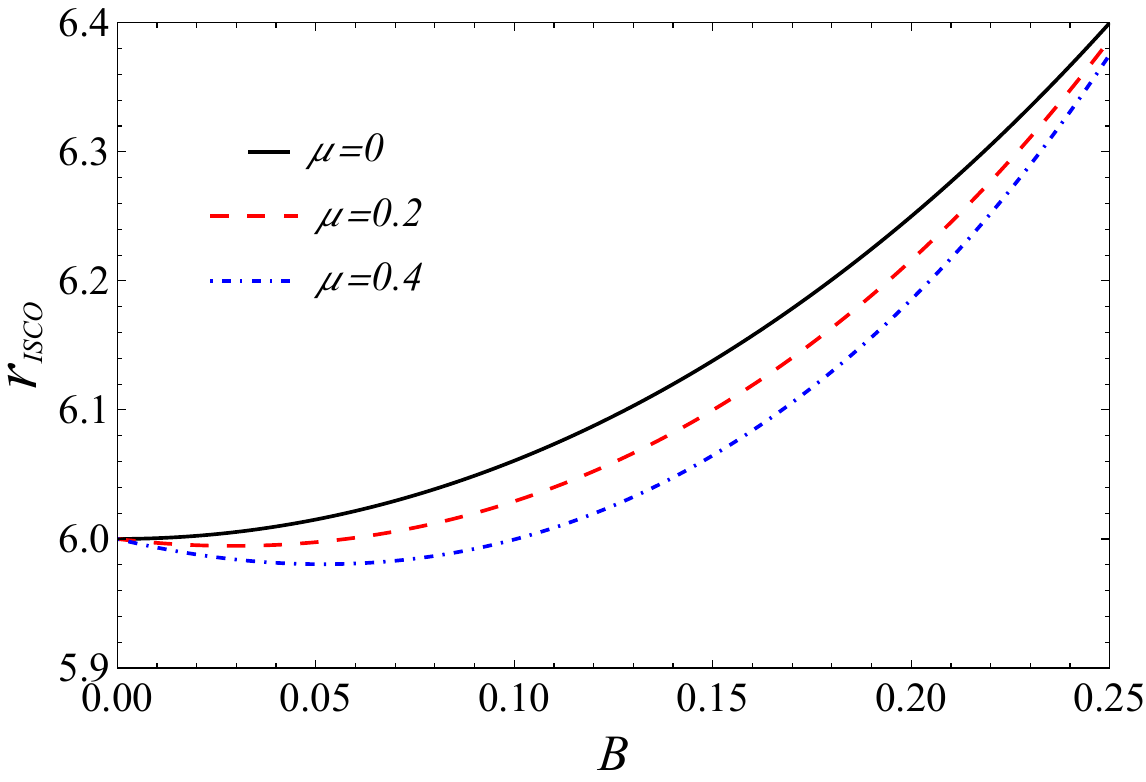}
\includegraphics[width=0.3\textwidth]{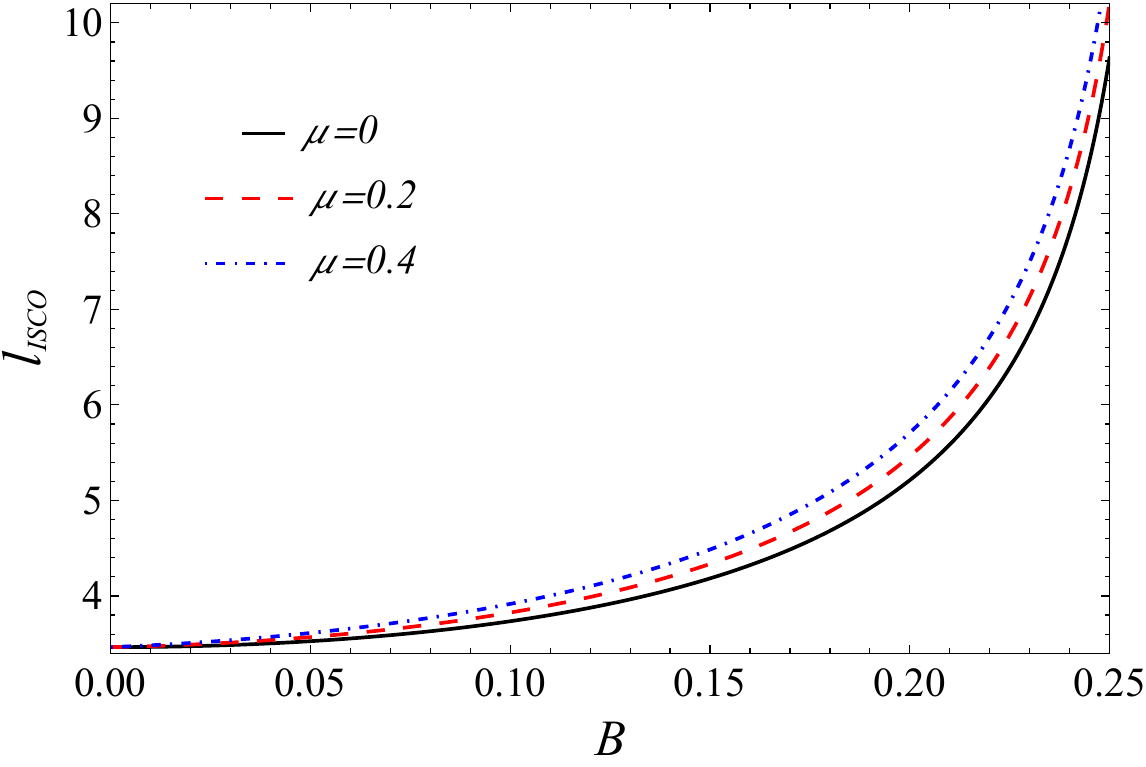}
\includegraphics[width=0.3\textwidth]{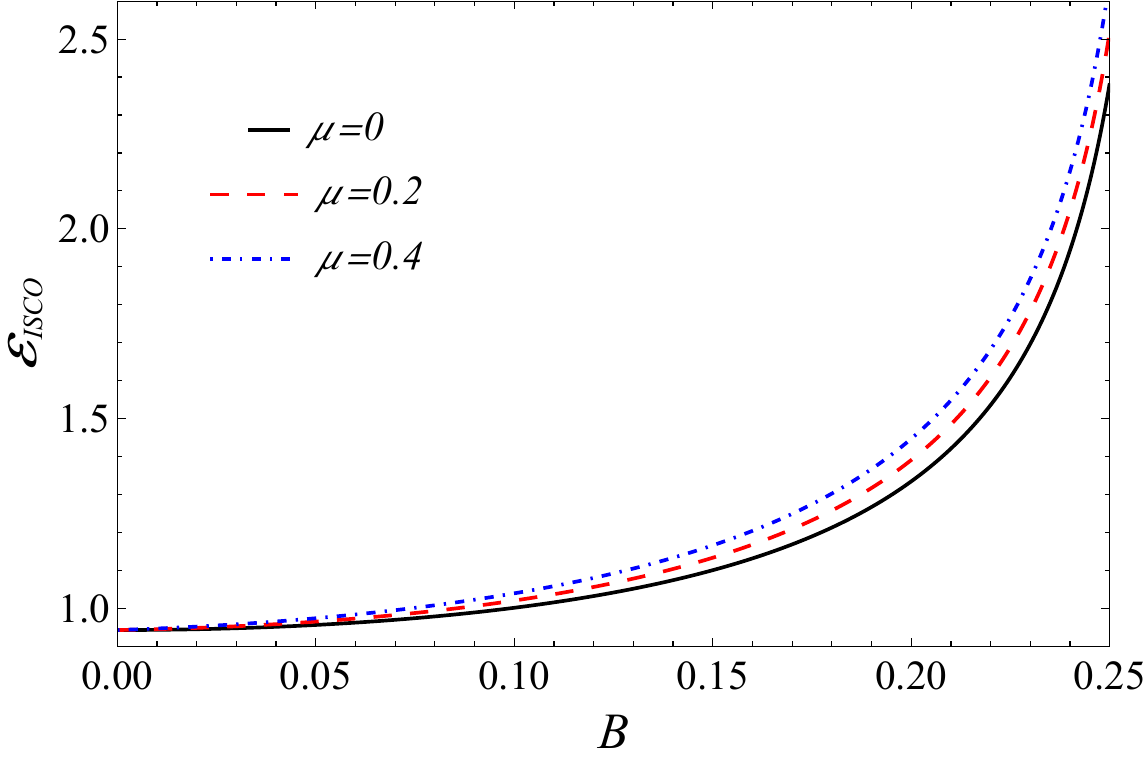}
\caption{Dependence of the ISCO radius $r_{\text{ISCO}}$, ISCO specific angular momentum $l_{\text{ISCO}}$, and the ISCO specific energy
$E_{\text{ISCO}}$ on the magnetic field $B$.\label{Fig.ISCO}}
\end{figure*}
\begin{figure*}[ht!]
\includegraphics[width=0.4\textwidth]{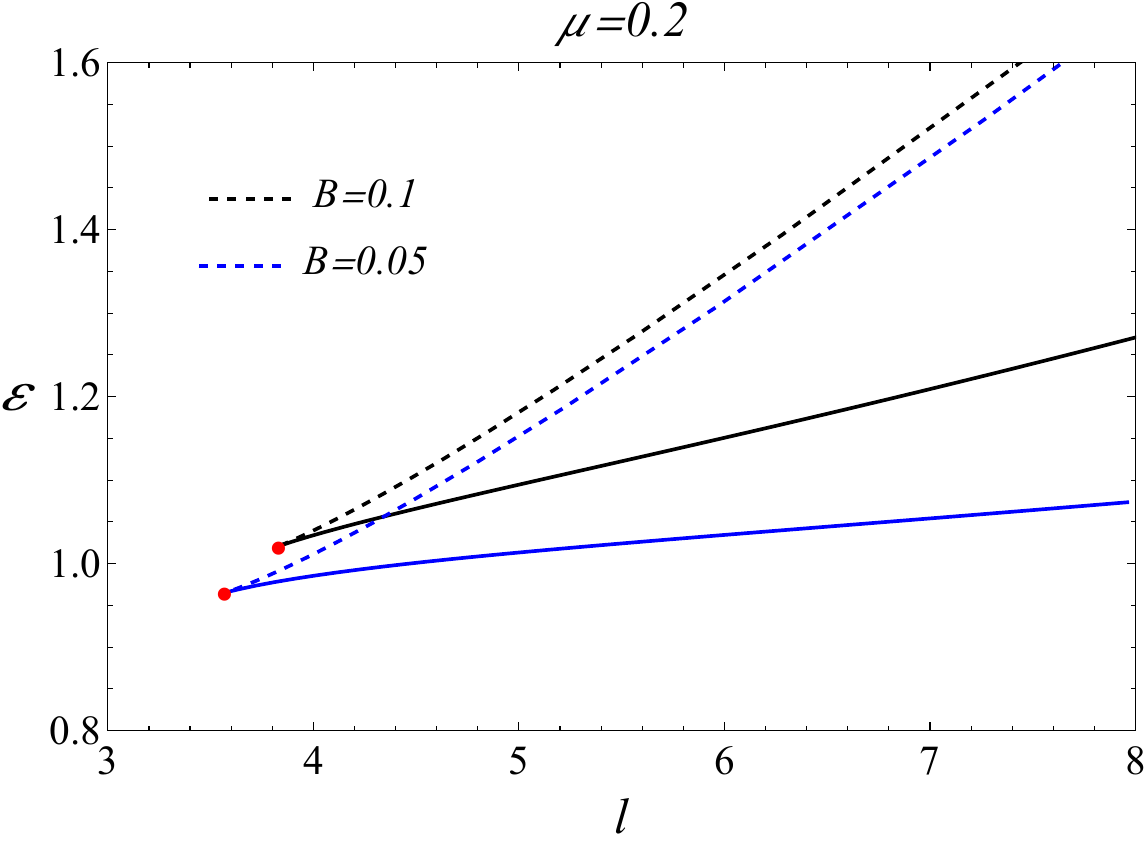}
\caption{Specific energy $\mathcal{E}$ and angular momentum $l$ for circular orbits at fixed value of the magnetic dipole momentum $\mu$. Here, stable circular orbits are depicted by solid lines, and unstable circular orbits are represented by dashed lines.
\label{fig. E l}}
\end{figure*}

Subsequently, we can find differential equation to govern the motion of the magnetized particles in the vicinity of the Schwarzschild BR using Eqs.~(\ref{eq.energy},\ref{eq.angular momen.}) as:
\begin{eqnarray}\label{eq.dr}
    \left(\frac{dr}{d\phi}\right)^2=\frac{r^4}{l^2} \big[\mathcal{E}^2-\mathcal{V}_1\big]\,.
\end{eqnarray}
The specific energy $\mathcal{E}$ and specific angular momentum $l$ of the magnetized particles can be found using conditions $\mathcal{E}^2=V_{eff}$ and $\partial_rV_{eff}=0$:
\begin{eqnarray}\label{eq.l E}
    l^2=\frac{r^2 M}{\mathcal{Y}}\bigg(1 +2\beta f(r) \bigg)\bigg(1 +\beta f(r) \bigg)\vert_{r=r_0}\,,
\\\mathcal{E}^2=\frac{\bigg(1 +\beta f(r)\bigg) f(r)^3}{\mathcal{Y}}\bigg( f(r) r + \beta \mathcal{Z} \bigg)\vert_{r=r_0}\,,
\end{eqnarray}
\begin{equation}
f(r) = \sqrt{1-B^2 M^2-\frac{2M}{r}}\,,
\end{equation}
\begin{equation}
N = 1+B^2 r^2\,,
\end{equation}
where we assumed that particle making circular orbits at the radius $r=r_0$ and new variables are given in Appendix~\ref{sec.app}.

Then we have depicted radial dependence of the effective potential for different values of the $\mu$ and $B$ in Fig.~\ref{Fig.effective}. The effective potential (\ref{eq.Veff1}) has two extreme points which correspond to the stable and unstable circular orbits. One can see from Fig.~\ref{Fig.effective} that increasing the values of the both parameters magnetic field $B$ and magnetic moment of the particle $\mu$ cause growing the value of the unstable circular orbits. However, enlarging the values of the  $B$ and $\mu$ leads decreasing the radii of the stable circular orbits.

Finding innermost stable circular orbits (ISCO) is important task to analyze the motion of the particles around BH which can be determined by the conditions \cite{Uktamov:2025bth,UktamjonUktamov:2025emm,Xamidov:2025hrj,Khan:2024jez}:
\begin{eqnarray}\label{eq.ISCO}
    V_{eff}(r_{ISCO})=\mathcal{E}_{ISCO}\,,\,\,\partial_rV_{eff}=\partial_r^2V_{eff}=0\vert_{r=r_{ISCO}}\,.
\end{eqnarray}
Conditions given in Eq.~(\ref{eq.ISCO}) paves a way to find ISCO radius $r_{ISCO}$, specific energy $\mathcal{E}_{ISCO}$ and specific angular momentum $l_{ISCO}$ at the ISCO, so we show dependence of the ISCO parameters on the magnetic field $B$ in Fig.~\ref{Fig.ISCO}. One can see from plots in Fig.(\ref{Fig.ISCO}) that increasing magnetic field $B$ and magnetic moment $\mu$ cause growing the values of the ISCO parameters $r_{ISCO}$, $\mathcal{E}_{ISCO}$, $l_{ISCO}$. 

Also, to be more informative we plot  region of parameter space where magnetized particles are permitted to have circular orbits in Fig.~\ref{fig. E l}. Red points in Fig.(\ref{fig. E l}) correspond to the specific energy $\mathcal{E}_{ISCO}$ and specific angular momentum $l_{ISCO}$ of the particle at ISCO.

\begin{figure*}[ht!]
\centering
\includegraphics[width=0.8\textwidth]{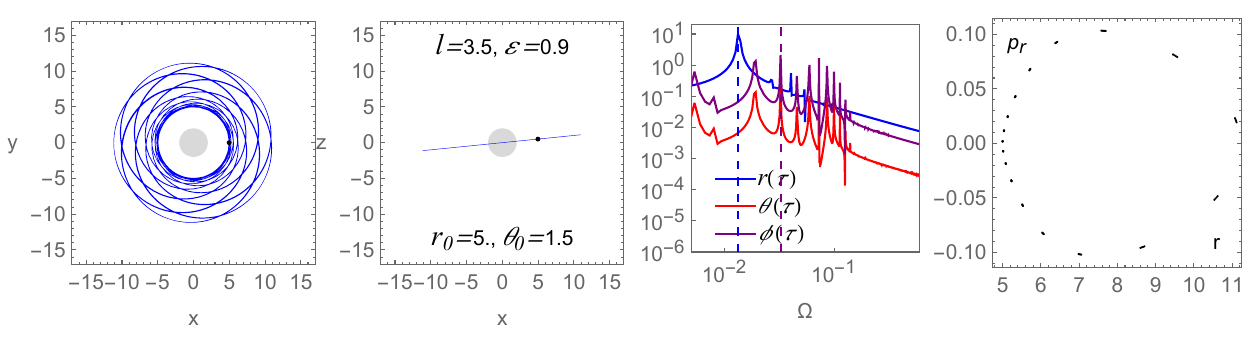}
\includegraphics[width=0.8\textwidth]{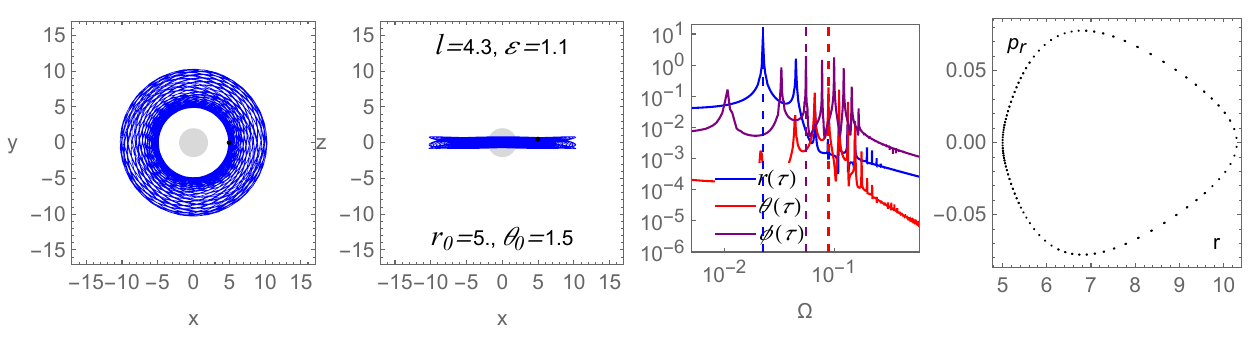}
\includegraphics[width=0.8\textwidth]{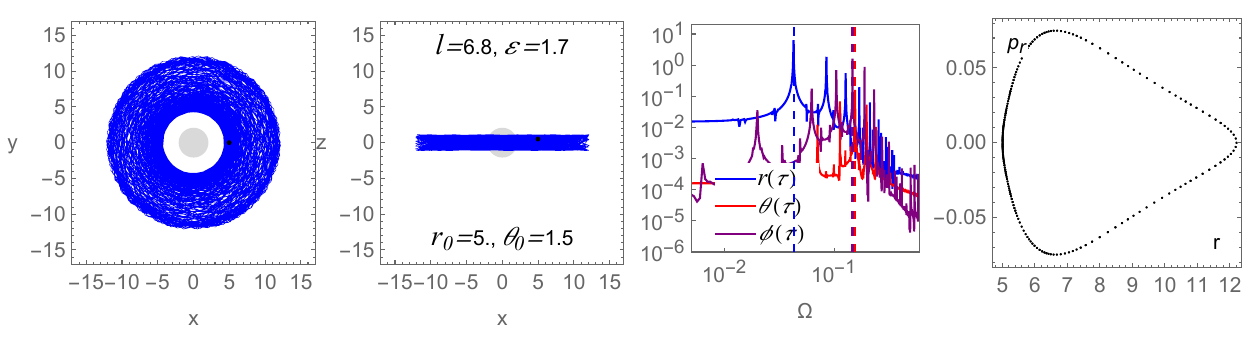}
\caption{\label{fig:Trajectory_for_magnetized} The trajectory of the magnetized particles (first and second column). Power Spectral Density for three $r(\tau)$ (blue), $\theta(\tau)$ (red), $\phi(\tau)$ (purple) coordinates along the particle’s trajectory (third column) and corresponding PS (last column) for different values of the magnetic field $B$ (from above row: $B=0, 0.1, 0.2$), here we take the value of the magnetic moment as $\mu/m=1$. [See Sec. 
\ref{sec:trajectory_dynamics_interpretation} for details].} 
\end{figure*}

Additionally, the trajectory of the magnetized particles in the vicinity of the Schwarzschild-BR BH is shown for different values of the magnetic field $B$ in Fig.~\ref{fig:Trajectory_for_magnetized} (first and second column). To analyze the chaotic behavior of the magnetized particle Poincaré section (PS) is given in the last column of the Fig.~\ref{fig:Trajectory_for_magnetized}~\cite{Kolos:2023oii}. One can conclude from plots in Fig.~\ref{fig:Trajectory_for_magnetized} that growing the value of the magnetic field $B$ regularize the motion of the of the magnetized particles in the vicinity of the Schwarzschild-BR BH (detailed explanation is given later).

\subsection{The Epicyclic Frequencies of the magnetized particles in Schwarzschild Berotti-Robinson black hole}\label{sec:Epicyclic_Frequencies_for_Bparticles_in_SBR_BH}
Now our aim is to find orbital frequency of the magnetized particles in the vicinity of the Schwarzschild BR BH, we start with the four-velocity normalization condition $u_\mu u^\mu=-1$:
\begin{eqnarray}\label{eq.normal}
g_{rr} \dot{r}^2 + g_{\theta\theta} \dot{\theta}^2 + V(r,\theta) = 0,
\end{eqnarray}
in which 
\begin{eqnarray}
V(r,\theta)= 1+\frac{\mathcal{E}^2g_{\phi\phi}+l^2g_{tt}}{g_{tt}g_{\phi\phi}\left(1-\beta\frac{\sin{\theta}}{R}\sqrt{PQ}\right)}\,,
\end{eqnarray}
here we use Eqs.~(\ref{eq.energy},\ref{eq.angular momen.}). We can expand $V(r,\theta)$ considering small perturbations($r_0 ,\theta_0$) near a circular orbit~\cite{2022Univ....8..507T}:
\begin{eqnarray}
r = r_0 + \delta r, \quad \theta = \theta_0 + \delta\theta\,,
\end{eqnarray}
as:
\begin{eqnarray}\label{eq.V}
    V(r,\theta)\approx\frac{1}{2}\partial_r^2V(r,\theta)\vert_{r_0,\theta_0}\delta r^2+\frac{1}{2}\partial_\theta^2V(r,\theta)\vert_{r_0,\theta_0}\delta\theta^2,
\end{eqnarray}
here we take conditions $V(r_0,\theta_0)=\partial_rV(r,\theta)=\partial_\theta V(r,\theta)=0$ into account~\cite{2022Univ....8..507T} as limiting condition of the constant of the motion and stability conditions must be fulfilled.

Then two independent harmonic oscillator equations can be derived using Eq.~(\ref{eq.normal}):
\begin{eqnarray}
\frac{d^2}{dt^2} \delta r + \Omega_r^2 \delta r = 0, \quad \frac{d^2}{dt^2} \delta \theta + \Omega_\theta^2 \delta \theta = 0,
\end{eqnarray}
where,
\begin{eqnarray}\label{eq.frequency}
\Omega_r^2 = \frac{1}{2g_{rr}\dot{t}^2} \partial_r^2V(r,\theta), \quad \Omega_\theta^2 = \frac{1}{2g_{\theta\theta}\dot{t}^2}  \partial_\theta^2V(r,\theta).
\end{eqnarray}

Finally, the explicit expressions for the epicyclic frequencies of the magnetized particles in Schwarzschild BH take the form:
\begin{widetext}
\begin{eqnarray}\label{eq.Omega r m}
 &\Omega_r^2=\frac{\Psi^2(r,\theta)}{2g_{rr}}\left[g_{tt}^2\partial_r^2\left(\frac{1}{g_{tt}\Psi(r,\theta)}\right)+g_{\phi\phi}^2\Omega_K^2\partial_r^2\left(\frac{1}{g_{\phi\phi}\Psi(r,\theta)}\right)\right],\\\label{Omega theta m}
 &\Omega_\theta^2=\frac{\Psi^2(r,\theta)}{2g_{\theta\theta}}\left[g_{tt}^2\partial_\theta^2\left(\frac{1}{g_{tt}\Psi(r,\theta)}\right)+g_{\phi\phi}^2\Omega_K^2\partial_\theta^2\left(\frac{1}{g_{\phi\phi}\Psi(r,\theta)}\right)\right],
\end{eqnarray}
\end{widetext}
in which $\Psi(r,\theta)=1-\beta\frac{\sin{\theta}}{\Omega}\sqrt{PQ}$ and $\Omega_K=-b{g_{tt}}/{g_{\phi\phi}}$ is the Keplerian frequency with the impact parameter $b={l}/{\mathcal{E}}$.

Then, the frequencies $\nu_i$ of the magnetized particles can be converted into physical units \cite{Kolos:2023oii}:
\begin{eqnarray}\label{eq.nu}
    \nu_i=\frac{1}{2\pi}\frac{c^3}{GM}\Omega_i\,\,\,\,(i=r,\theta,K).
\end{eqnarray}

\subsection{ Trajectory dynamics interpretation for magnetized particles}\label{sec:trajectory_dynamics_interpretation}

We study the trajectory dynamics, and we provide the profiles of their parametrisation as follows.

Fig. \ref{fig:Trajectory_for_magnetized} presents trajectory data for magnetized particles in a \(3 \times 4\) panel matrix. Each row corresponds to a set of 4 panel, and each row corresponds to a different parametrisation, i.e. either the magnetic field strength, $B$ change, or particle charge, $q$ change. 

The four columns display:

\begin{itemize}
    \item \textbf{Column 1:} \(xy\)-projection of the trajectory.
    \item \textbf{Column 2:} \(xz\)-projection, along with the specific angular momentum \(l\), specific energy \(\epsilon\), initial radial coordinate \(r_0\), and initial angle \(\theta_0=\frac{\pi}{2}+\delta \theta \).
    \item \textbf{Column 3:} Power spectral density (PSD) for the spherical coordinates \(r(\tau)\), \(\theta(\tau)\), and \(\phi(\tau)\).
    \item \textbf{Column 4:} Phase space of the radial component.
\end{itemize}

In Fig.~\ref{fig:Trajectory_for_magnetized}, each row corresponds to a panel with a different  magnetic field strength \(B=0\) (first row), \(B=0.1\) (second row), and \(B=0.2\) (third row). This case corresponds to the initial radial coordinate \(r_0=5.0\), and initial angle \(\theta_0=1.5\).

As the magnetic field increases, both \(l\) and \(\epsilon\) increase accordingly: \((l, \epsilon) = (3.5, 0.9)\) for \(B=0\), \((4.3, 1.1)\) for \(B=0.1\), and \((6.8, 1.7)\) for \(B=0.2\).

For each column, of Fig.~\ref{fig:Trajectory_for_magnetized}, we observe the following.

\paragraph{Column 1:}
In all cases, the particle follows a circular, epicyclic trajectory. The paths become more densely packed as the magnetic field increases, as expected.

\paragraph{Column 2:}
In the first row (\(B=0\)), the trajectory shows a slight tilt in the \(xz\)-plane. This is due to the initial condition \(\theta_0 = \frac{\pi}{2} + \delta\theta_0\), i.e., a small deviation from the equatorial plane. In the second and third rows, the trajectory stabilizes horizontally, indicating that the magnetic field confines the particle motion near the equatorial plane, suppressing vertical excursions.

\paragraph{Column 3:}
The Power Spectral Density (PSD) for the spherical coordinates \(r(\tau)\), \(\theta(\tau)\), and \(\phi(\tau)\) is presented. As the magnetic field increases (from top to bottom row), the main frequency peak shifts to higher values, indicating the increasing influence of the Lorentz force on the characteristic timescales of motion. For regular trajectories oscillating around effective potential minima (e.g., the \(B=0\) case), the PSD shows distinct main peaks corresponding to the fundamental frequencies calculated analytically, along with their higher harmonics (overtones). As the magnetic field increases to intermediate values (\(B=0.1, 0.2\)), the high-frequency region transitions from smooth to increasingly noisy, reflecting the onset of chaotic dynamics. We also observe that the azimuthal motion \(\phi(\tau)\) exhibits the strongest high-frequency components, consistent with the higher curvature of this degree of freedom.

\paragraph{Column 4:}
The radial phase space (PS), also known as the Poincaré section (PS) shows chaotic behavior in the absence of a magnetic field. As the magnetic field increases, the phase space becomes more regular, indicating a stabilizing effect due to the magnetic field.

\subsection{Harmonic oscillation frequencies}\label{sec:Harmonic_oscillation_frequencies}

 \begin{figure*}[ht!]\centering
\includegraphics[width=0.3\textwidth]{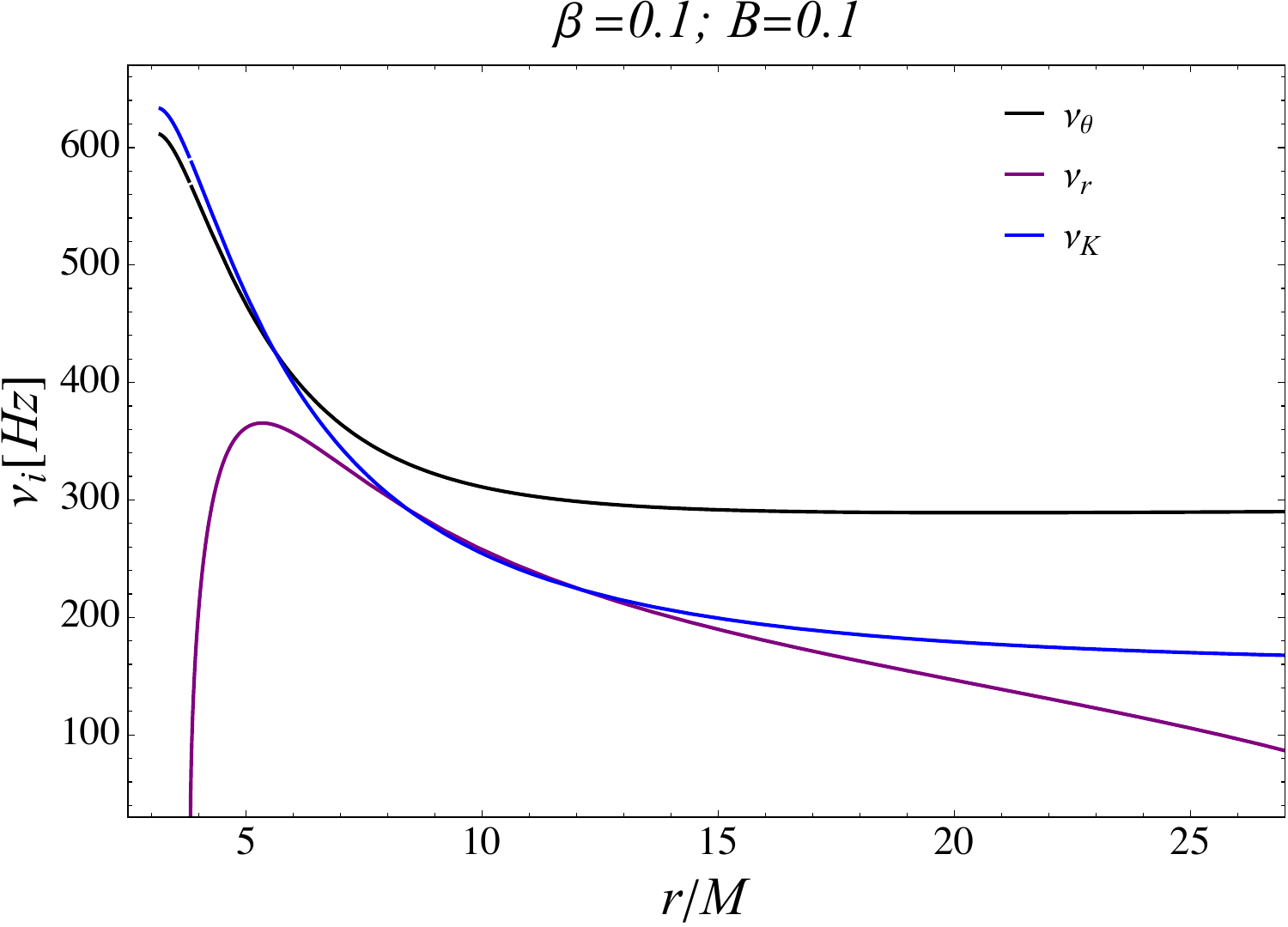}
\includegraphics[width=0.3\textwidth]{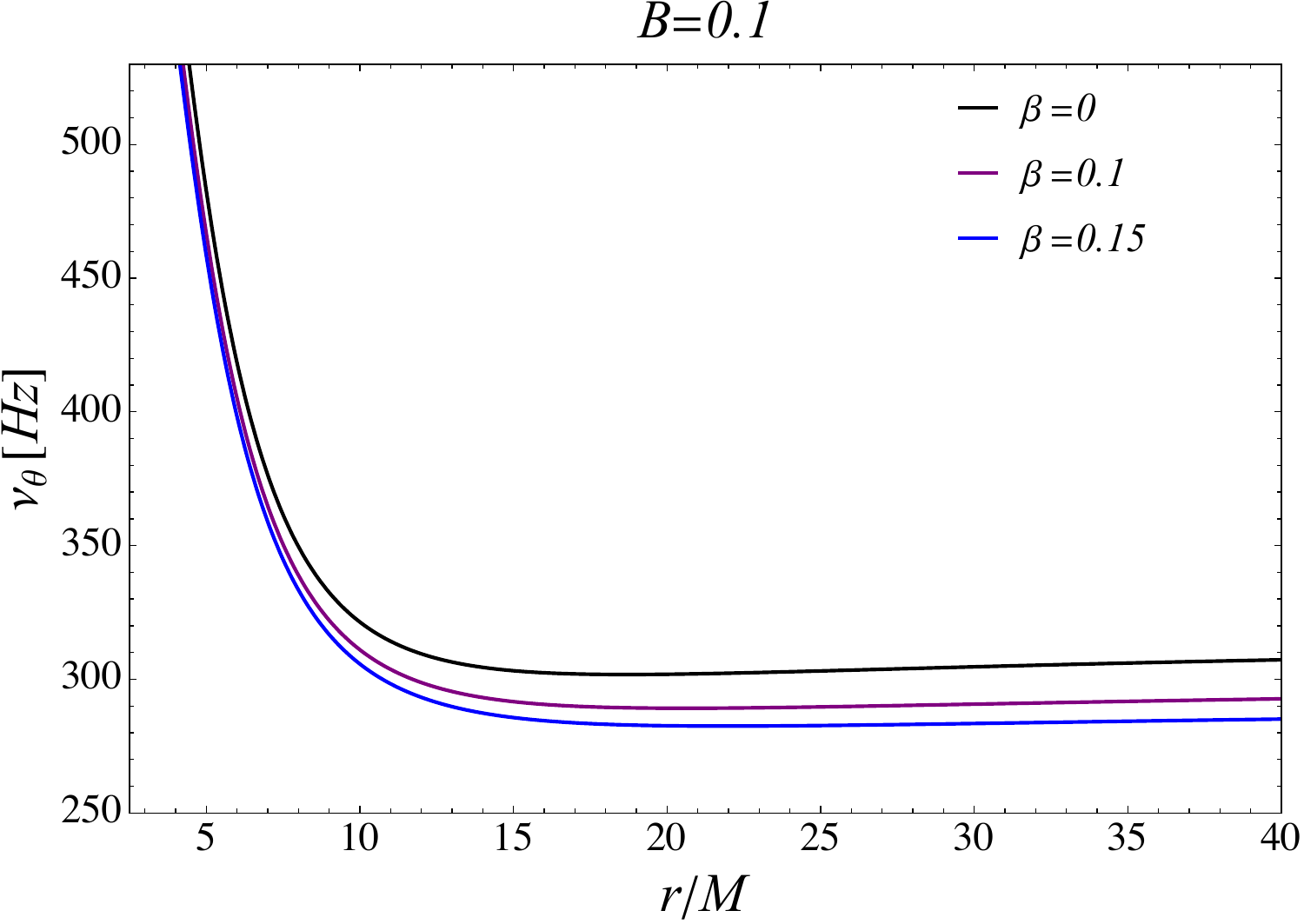}
\includegraphics[width=0.3\textwidth]{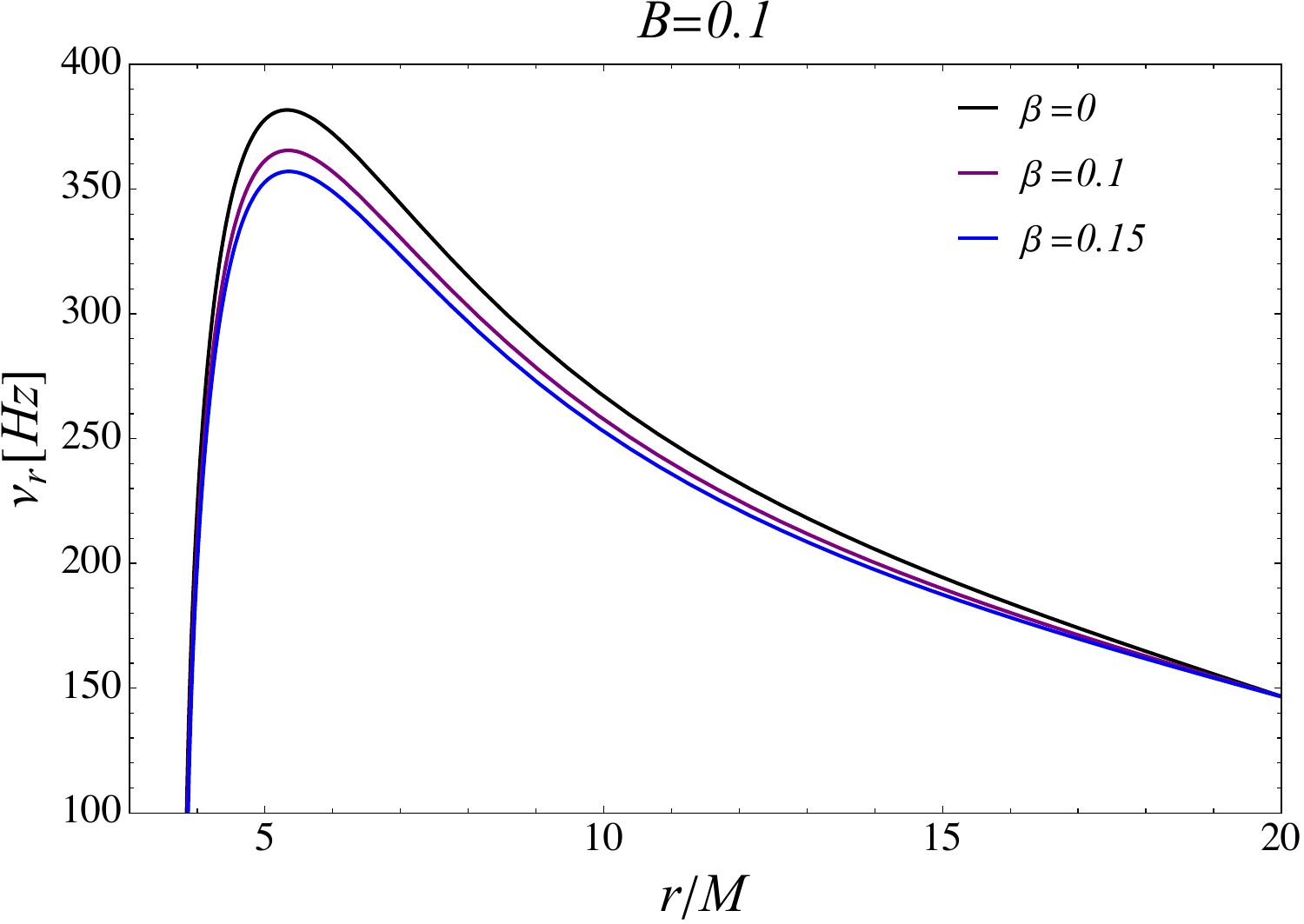}
\includegraphics[width=0.3\textwidth]{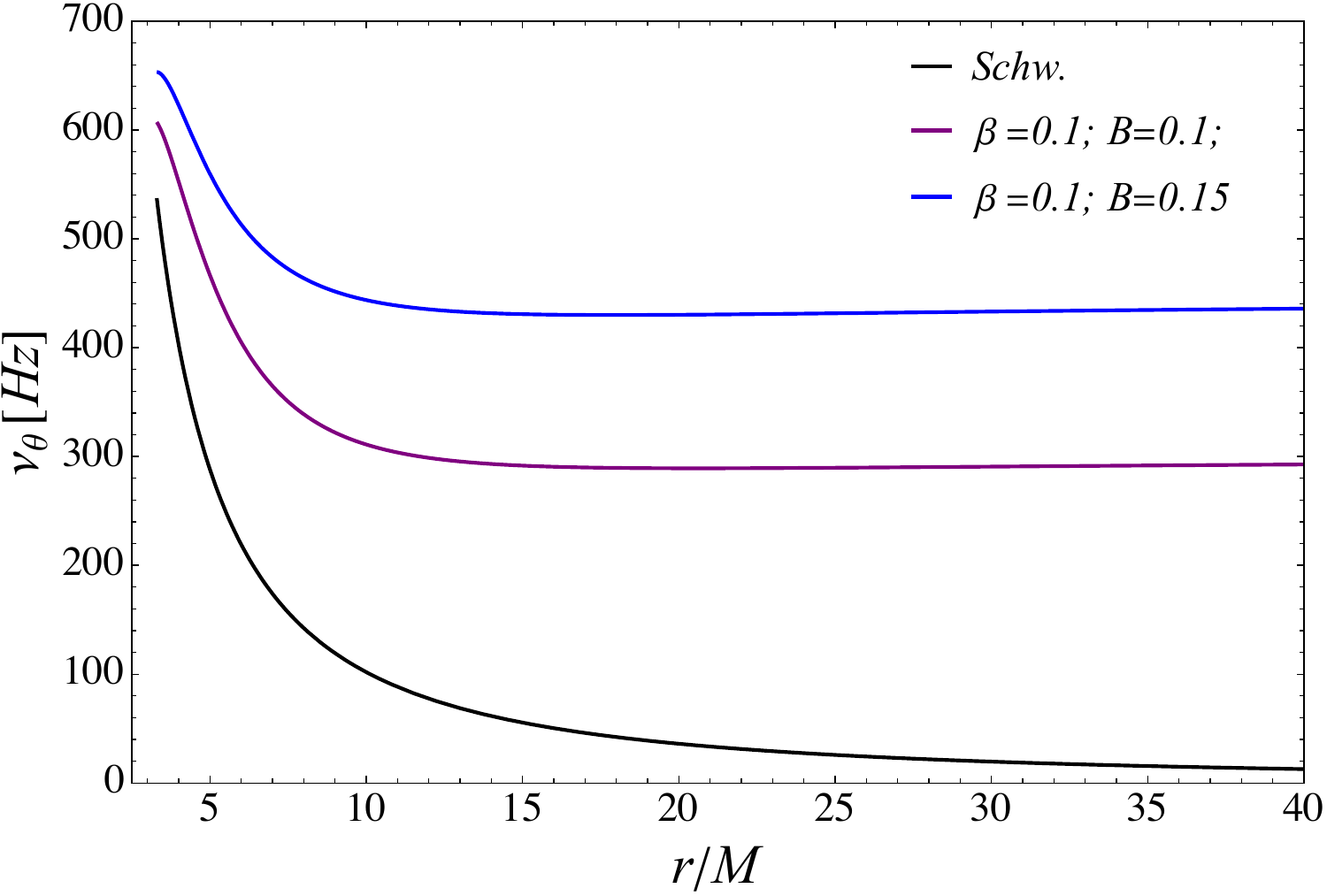}
\includegraphics[width=0.3\textwidth]{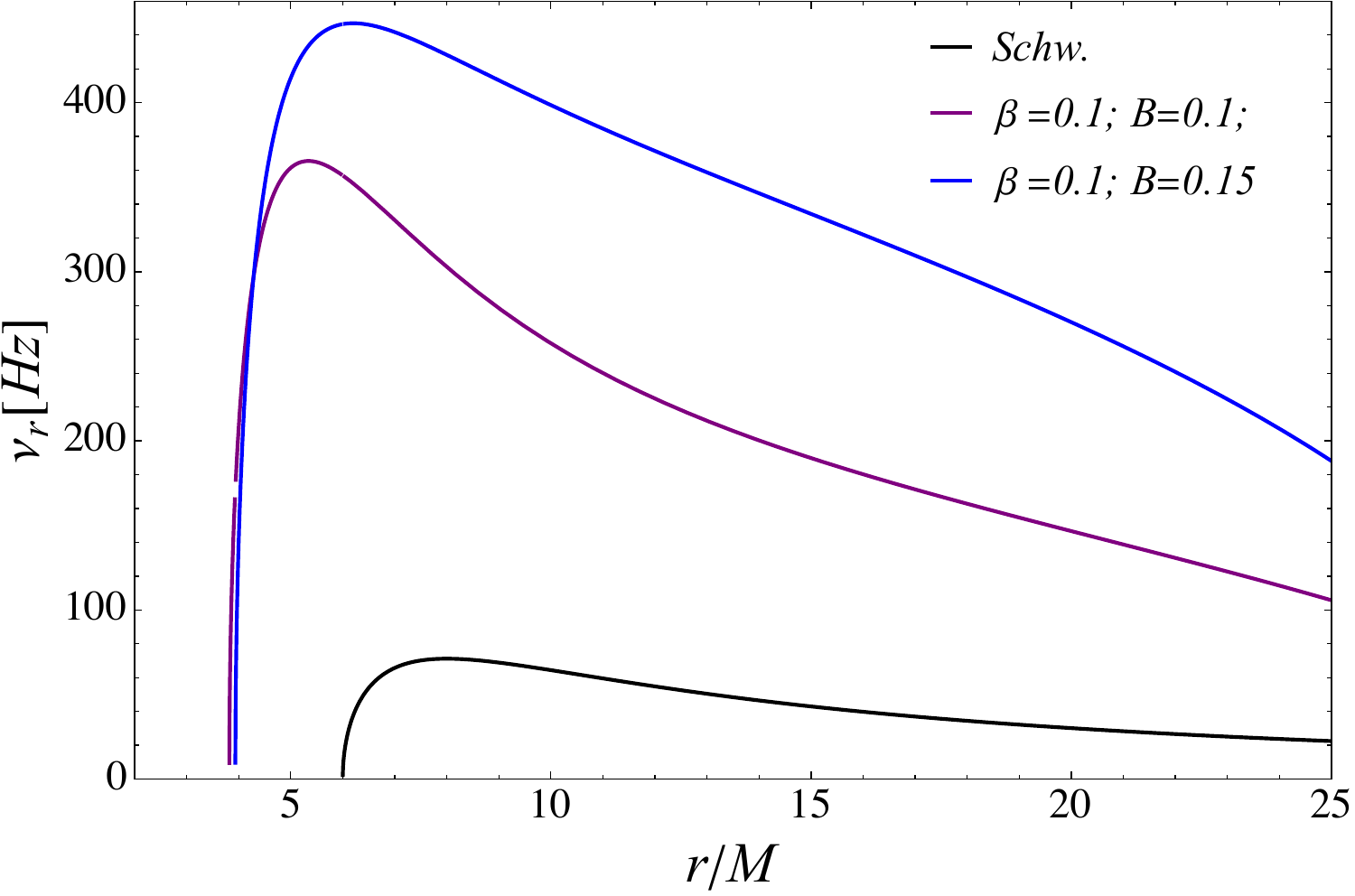}
\caption{\label{Fig.frequency magnetized}An investigation into how the radial profiles of the small harmonic oscillation frequencies ($\nu_r$, $\nu_\theta$, and $\nu_K$) of magnetized particles are modified by the parameters $B$ and $\beta$ in the equatorial plane of a Schwarzschild-BR black BH mass $M=10M_\odot$.}
\end{figure*}

Subsequently, we have plotted radial dependence of the parameters $\nu_\theta$, $\nu_r$, $\nu_K$ for different values of the parameters $B$, $\beta$ in Fig.~\ref{Fig.frequency magnetized}. One can see from plots in Fig.~\ref{Fig.frequency magnetized} that the values of the harmonic oscillation frequencies ($\nu_\theta$, $\nu_r$, $\nu_K$) decreasing with increasing the values of the $r$. Also, it is seen from Fig.~\ref{Fig.frequency magnetized} that presence of the magnetic field leads growing the values of the frequencies.

Additionally, in the third column of Fig.~\ref{fig:Trajectory_for_magnetized}, we show how the magnetic field $B$ affects the  the Fourier spectra, i.e. PSD obtained from the magnetized particle trajectories. For particles with magnetic dipole momentum on nearly circular orbits, the dominant peaks in their power spectra correspond to the fundamental frequencies defined in Eqs.~(\ref{eq.frequency}). These plots in Fig.~\ref{fig:Trajectory_for_magnetized} illustrates again, that increasing the values of the magnetic field $B$ of the Schwarzschild-BR BH leads growing the values of the $\nu_\theta$, $\nu_r$, $\nu_K$.

\section{The motion of the electrically charged particles.}\label{sec:Motion_of_electrically_charged_particles}
Now we investigate the dynamics of the electrically charged particles around Schwarzschild-BR BH.  
For charged particles moving in the spacetime (\ref{eq.metric}), the Lagrangian takes the form:
\begin{eqnarray}\label{eq.charged lagrangian}
    \mathscr{L}=\frac{1}{2}mg_{\mu\nu}u^\mu u^\nu+q u^\mu A_\mu,
\end{eqnarray}
which paves a way to find four-momentum as:
\begin{eqnarray}\label{eq.four momentum char.}
   p_\mu=\frac{\partial\mathscr{L}}{\partial\Dot{x}^\mu}=mg_{\mu\nu}u^\nu+qA_\mu. 
\end{eqnarray}
Subsequently, Hamilton-Jacobi equation for the electrically charged particle can be expressed as:
\begin{eqnarray}\label{eq.H-J for charged}
   g^{\mu\nu}\left(\frac{\partial S}{\partial x^\mu}+qA_\mu\right)\left(\frac{\partial S}{\partial x^\nu}+qA_\nu\right)=-m^2\,, 
\end{eqnarray}
where $q$ and $m$ are the electric charge and the mass of the test particle, respectively. 

\begin{figure*}[ht!]
\includegraphics[width=0.45\textwidth]{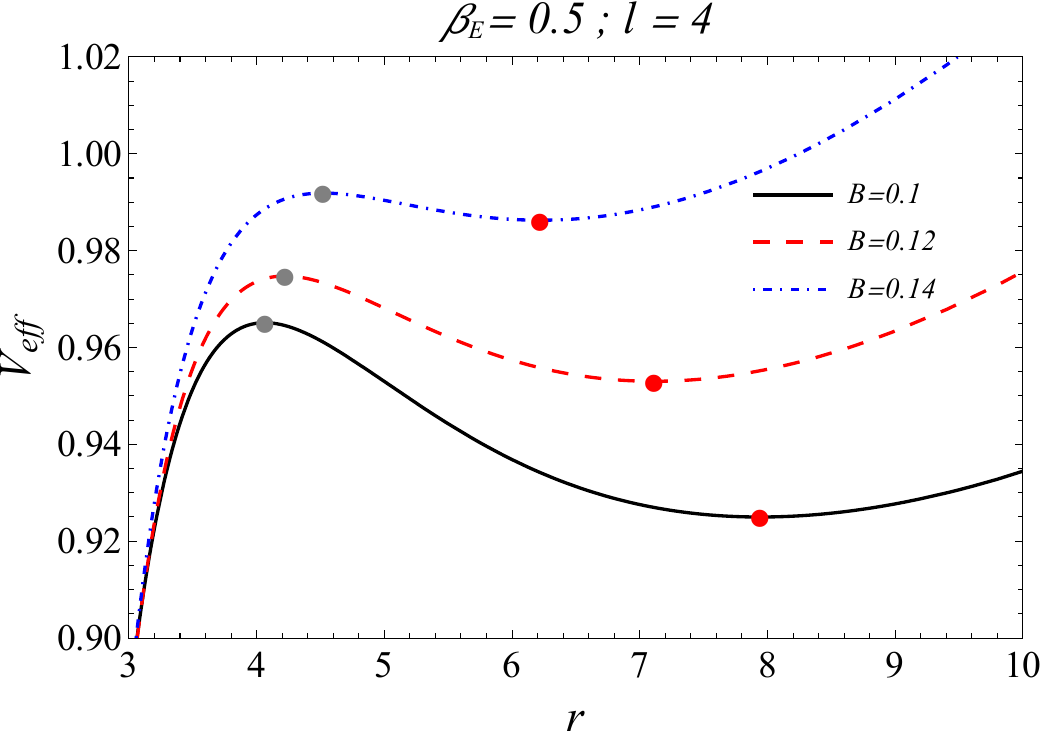}
\includegraphics[width=0.45\textwidth]{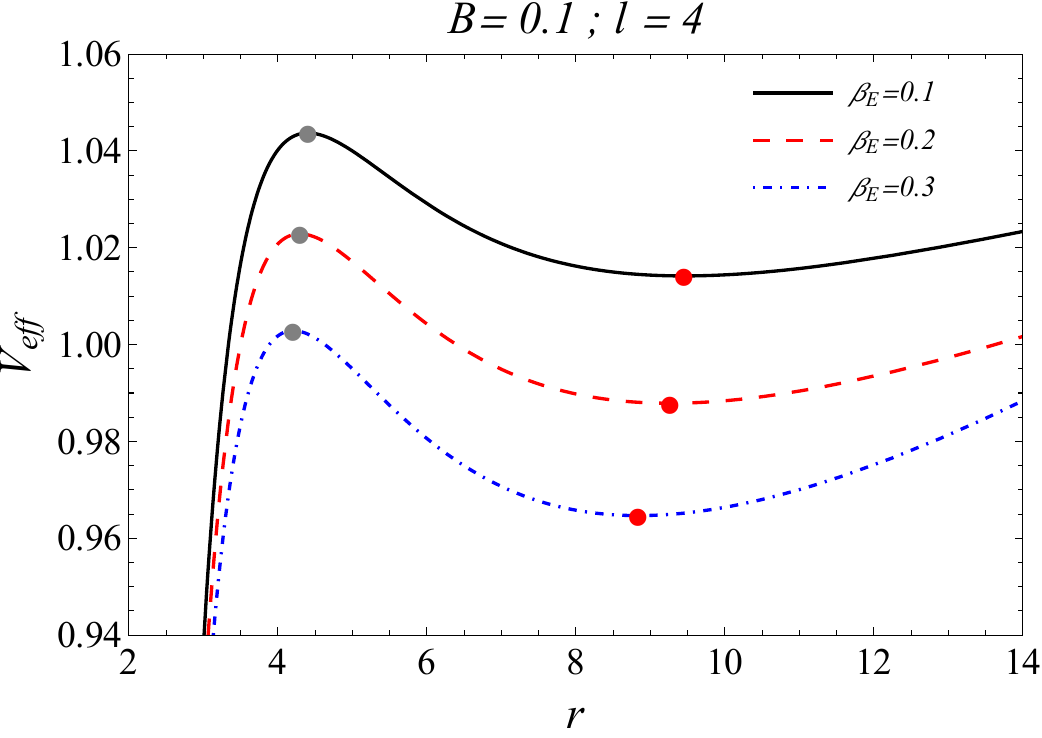}
\caption{The radial dependence of the effective potential $V_{eff}$ of the electrically charged particle for different values of the $B$ and $\beta_E$. \label{Fig.effective charg.}}
\end{figure*}

Again, considering the motion of the charged particle at the equatorial plane the equation of the motion can be found as:
\begin{subequations}\label{eq.motion of ch.}
    \begin{align}
        & \dot{t}=-\frac{\mathcal{E}}{g_{tt}}\,,\\
        &\dot{\phi}=\frac{l-\beta_EA_\phi}{g_{\phi\phi}}\,,\\
        & \dot{r}^2=R^4\left(\mathcal{E}^2-V_{eff}\right)\,,
    \end{align}
\end{subequations}
where $\beta_E=q/m$ and $V_{eff}$ is:
\begin{widetext}
\begin{eqnarray}\label{eq.Vefff cha.}
V_{eff} = \left(1-\frac{2M}{r}-B^2M^2\right)
\left\{1+r^{-2}(1+B^2r^2) \left[l+\frac{\beta_E}{B}\left(1-\frac{1}{\sqrt{1+B^2r^2}}\right)^2\right]^2\right\} \,,
\end{eqnarray}
\end{widetext}

or we can approximate Eq.(\ref{eq.Vefff cha.}) as:
\begin{widetext}
\begin{eqnarray}\label{eq.Vefff cha2.}
    V_{eff}=f_{Schw}(r)\left(1+\frac{l^2}{r^2}-Bl\beta_E\right)+B^2\left[f_{Schw}(r)\left(\frac{\beta_E^2r^2}{4}+l^2\right)-M^2(1+\frac{l^2}{r^2})\right]+\mathcal{O}(B^2)\,,
\end{eqnarray}
\end{widetext}
where $f_{Schw}(r)=1-\frac{2M}{r}$ is the lapse function for the Schwarzschild BH.
 Then we show how effective potential $V_{eff}$ from Eq. (\ref{eq.Vefff cha.}) depends on the $r$ for fixed values of the parameters $B$ and $\beta_E$ in Fig.~\ref{Fig.effective charg.}. Similarly, effective potential in Fig.~\ref{Fig.effective charg.} has two extreme points corresponding to the stable circular orbits (red points in Fig.~\ref{Fig.effective charg.}) and unstable circular orbits (gray points in Fig.~\ref{Fig.effective charg.}). One can notice from Fig.~\ref{Fig.effective charg.} that increasing the values of the both parameters $B$ and $\beta_E$ leads growing the values of the stable circular orbits while enlarging the values of the $B$ and $\beta_E$ causes decreasing the radii of the unstable circular orbits.

\begin{figure*}[ht!]
\includegraphics[width=0.3\textwidth]{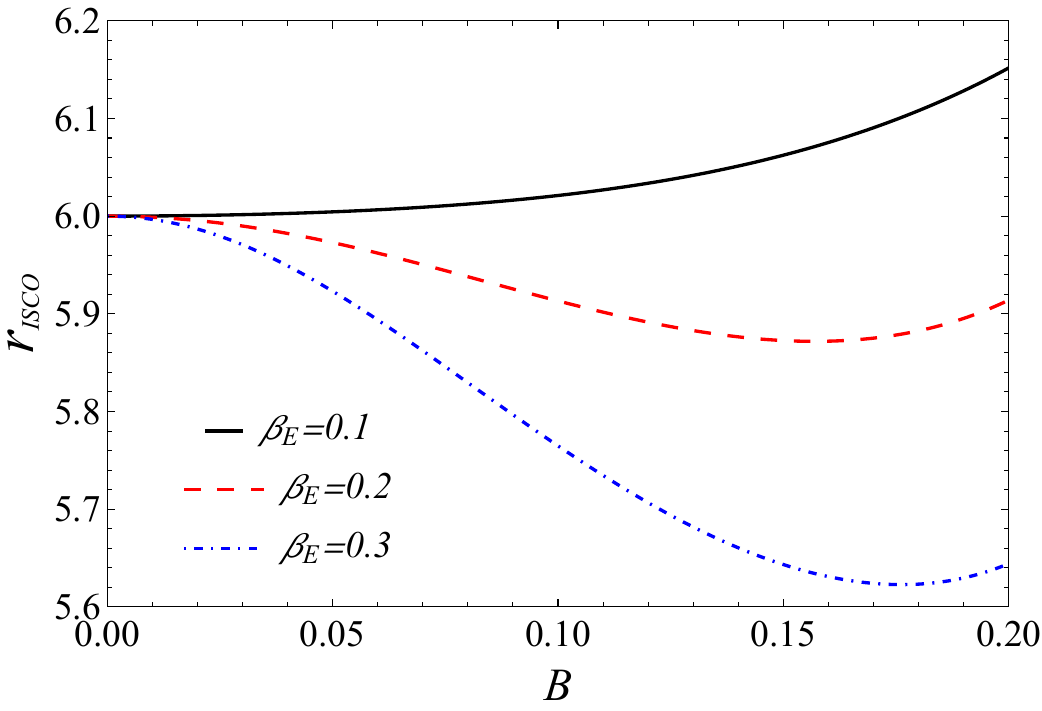}
\includegraphics[width=0.3\textwidth]{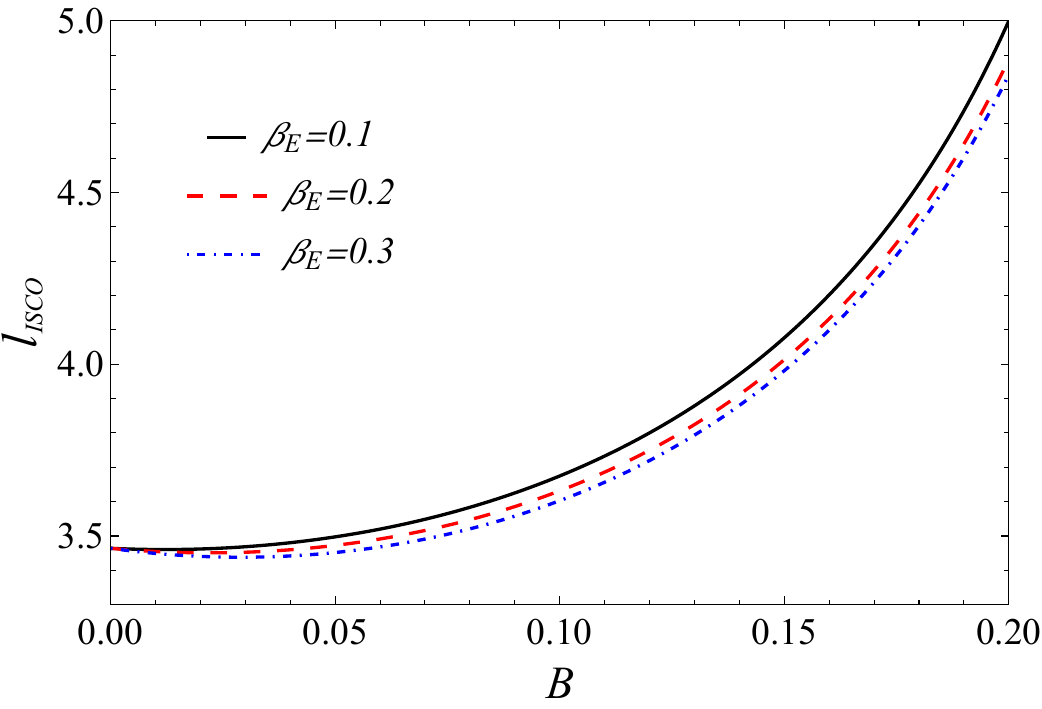}
\includegraphics[width=0.3\textwidth]{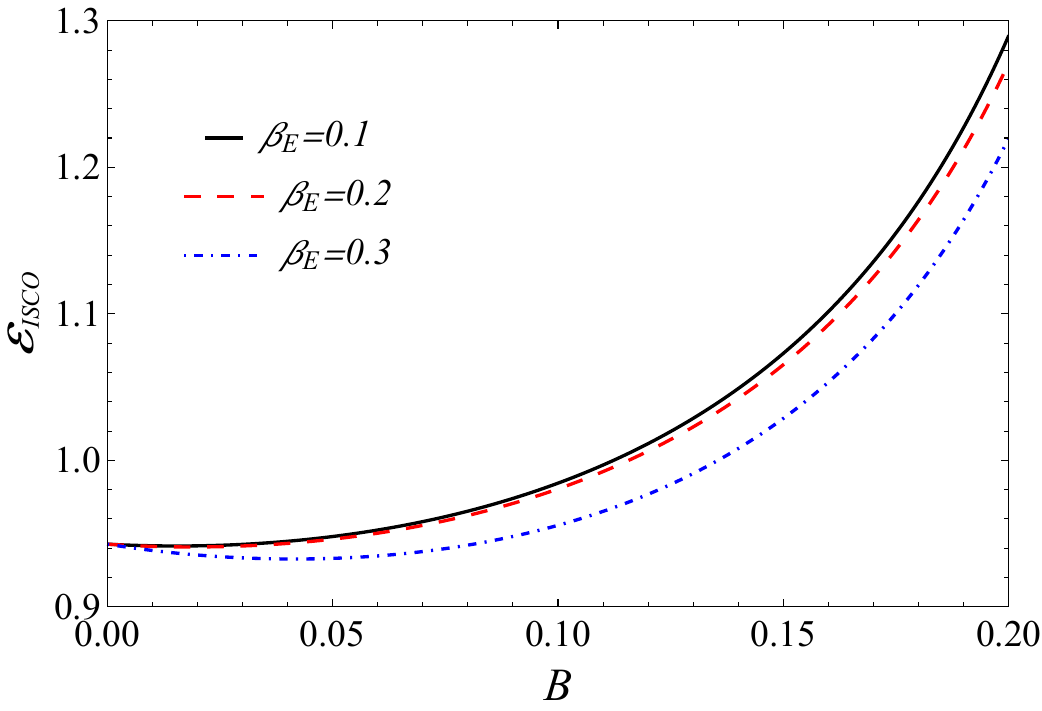}
\caption{Plots of the ISCO parameters $r_{ISCO}$, $l_{ISCO}$, and $\mathcal{E}_{ISCO}$ versus the magnetic field $B$, shown for different $\beta_E$ values.\label{Fig. ISCO ch.}}
\end{figure*}

Consistent with the results discussed above, now we analyze ISCO parameters for the charged particle in the vicinity of the Schwarzschild-BR BH:
\begin{eqnarray}\label{eq.ISCO char.}
 V_{eff}(r)=\mathcal{E}^2_{ISCO}\,,\,\,\,\,\partial_rV_{eff}=\partial_r^2V_{eff}=0\,,   
\end{eqnarray}
here $V_{eff}$ is given with Eq.~(\ref{eq.Vefff cha.}). We illustrate dependence of the ISCO parameters $r_{ISCO}$, $l_{ISCO}$, and $\mathcal{E}_{ISCO}$ on the magnetic field of the Schwarzschild-BR BH for fixed values of $\beta_E$ in Fig.~\ref{Fig. ISCO ch.}. It is clear from Fig.~\ref{Fig. ISCO ch.} that increasing the values of the $B$ yield growing the value of the all ISCO parameters $r_{ISCO}$, $l_{ISCO}$ and $\mathcal{E}_{ISCO}$ and the particle with higher electric charge $q$ has larger ISCO radius $r_{ISCO}$ compared to the particle with lower electric charge but high electric charge $q$ causes smaller values of the ISCO angular momentum $l_{ISCO}$ and ISCO specific energy $\mathcal{E}_{ISCO}$.

Furthermore, using Eqs.(\ref{eq.motion of ch.}) the differential equation to govern the motion of the electrically charged particles can be found:
\begin{eqnarray}\label{eq.drda}
    \left(\frac{dr}{d\phi}\right)^2=\frac{r^4}{\left[l-\frac{\beta_E}{B}\left(-1+\frac{1}{\sqrt{1+B^2r^2}}\right)\right]^2} \big[\mathcal{E}^2-\mathcal{V}_2\big]\,.
\end{eqnarray}

 The specific energy $\mathcal{E}$ and specific angular momentum $l$ of the electrically particles orbiting with circular radius $r=r_0$ can be found using conditions $\mathcal{E}^2=V_{eff}$ and $\partial_rV_{eff}=0$:
 
 \begin{eqnarray}\label{eq.angular momentum.ch}
     l&=&\frac{r^2}{2\mathcal{Y}}\left(\mathcal{A}-B M \beta_E\right)\vert_{r=r_0}\,,
\\\label{eq.energy ch.}
\mathcal{E}^2&=&\frac{1}{4 r \mathcal{Y}^2}\bigg(4\mathcal{Y}\mathcal{B}_1+\mathcal{A} \mathcal{B}_2-\mathcal{B}_3-\mathcal{B}_4\bigg)\vert_{r=r_0}\,.
\end{eqnarray}
 

\begin{figure*}[ht!]
\includegraphics[width=0.45\textwidth]{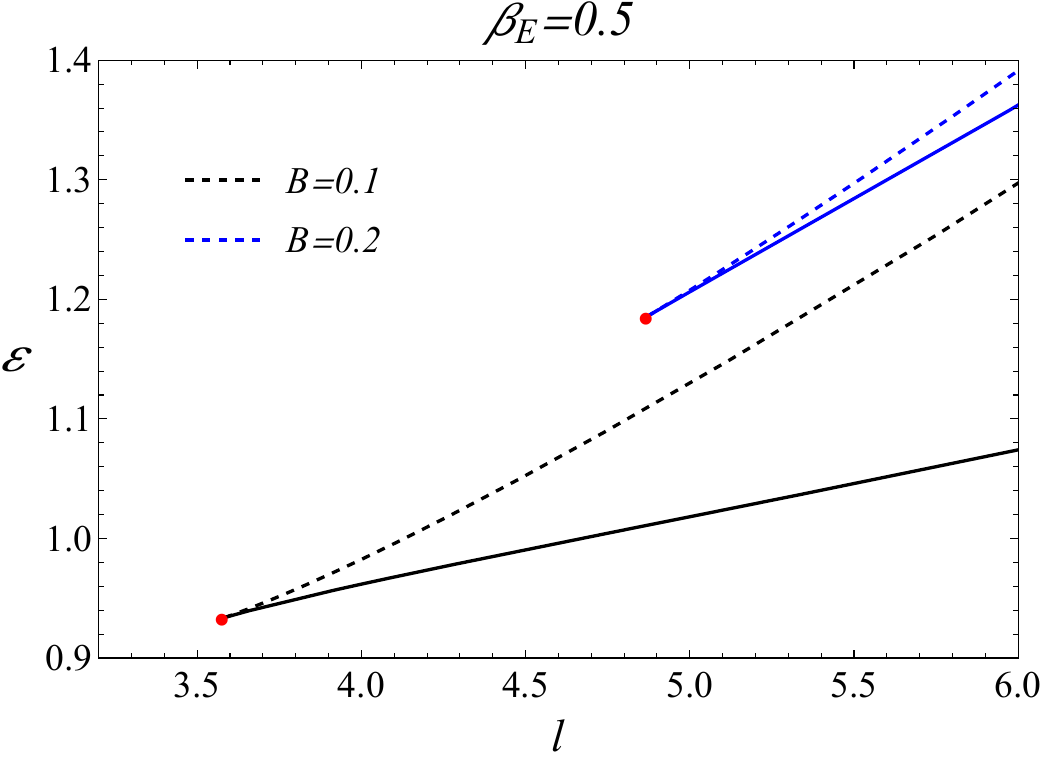}
\includegraphics[width=0.45\textwidth]{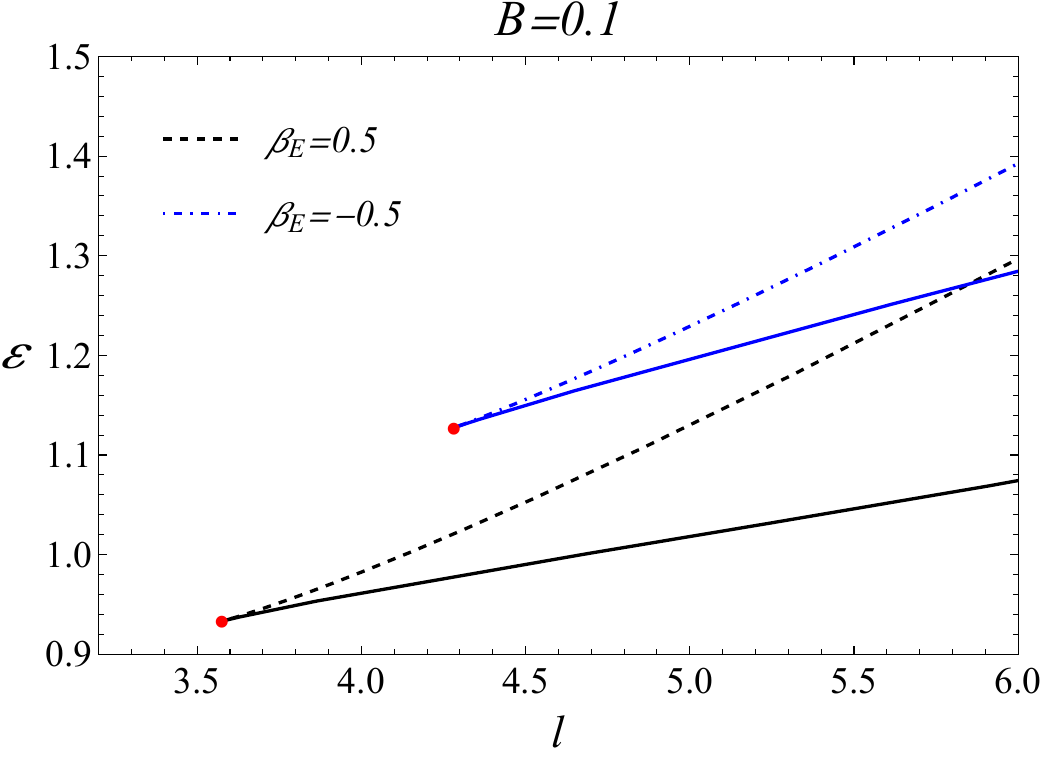}
\caption{Values of the specific energy $\mathcal{E}$ and specific angular momentum $l$ for circular orbits plotted for fixed values of the magnetic coupling parameter. Here, stable circular orbits are depicted by solid lines, and unstable circular orbits are represented by dashed lines.\label{Fig.l e space for charged}}
\end{figure*}

We also illustrate in Fig.~\ref{Fig.l e space for charged} the range of parameters that allow electrically charged particles to follow circular trajectories.

Additionally, we show how the parameter $B$ (Fig. \ref{Fig.trajectory ch.1}, first and second columns) and the electric charge $q$ (Fig. \ref{Fig.trajectory ch.2}, first and second columns) affect the trajectory of charged particles in the Schwarzschild-BR BH spacetime. To analyze the chaotic behavior of the magnetized particle, Poincaré section (PS) is given in the last column of Figs.~\ref{Fig.trajectory ch.1} and \ref{Fig.trajectory ch.2}. As demonstrated in Figs.~\ref{Fig.trajectory ch.1} and \ref{Fig.trajectory ch.2}, larger values of the magnetic field $B$ and electric charge $q$ result in more regular trajectories for charged particles around the Schwarzschild-BR BH. (detailed explanation is given in (\ref{sec:charged_particle_trajectory_dynamics_interpretation}))

\subsection{The Epicyclic frequencies of the electrically charged particles in Schwarzschild Berotti-Robinson black hole}

To find angular frequencies $\Omega_E$ of the electrically charged particles we assume motion of particles in circular orbit with four velocity $u^\mu=\Dot{t}(1,0,0,\Omega_E)$, so normalization condition $u_\mu u^\mu=-1$ enables~\cite{Turimov:2022evw}:
\begin{eqnarray}\label{eq.t ch}
 \Dot{t}=\frac{1}{\sqrt{-g_{tt}-\Omega_E^2g_{\phi\phi}}}\,.   
\end{eqnarray}
Also, we can write covariant Lorentz equation to govern the motion of the charged particle as:
\begin{eqnarray}\label{eq.Lorentz for ch.}
    \frac{du^\alpha}{d\lambda}+\Gamma_{\mu\nu}^\alpha u^\mu u^\nu=\beta_EF_\beta^\alpha u^\beta\,,
\end{eqnarray}
here $\lambda$ is the affine parameter and $\Gamma_{\mu\nu}^\alpha u^\mu$ is the Christoffel symbols.

Then solving Eq. (\ref{eq.Lorentz for ch.}) for radial and vertical motion paves a way to find expressions for the Larmor frequencies $\Omega_L$ which appears due to magnetic field and pure gravitational frequency $\Omega_G$ as:
\begin{eqnarray}\label{eq.Omega ch}
    \Omega_G^2=-\frac{g_{tt,r}}{g_{\phi\phi,r}}\,,\,\,\,\,\Omega_L=-\frac{2\beta_E}{g_{\phi\phi,\theta}}\frac{F_{\theta\phi}}{\Dot{t}}\,.
\end{eqnarray}
Following, angular velocity of the charged particle can be expressed as $\Omega_E=\Omega_G+\Omega_L$, so after putting Eq. (\ref{eq.t ch}) into Eq. (\ref{eq.Omega ch}) we have expression:
\begin{eqnarray}\label{eq.exp}
\Omega_L^2(1+\gamma)+2\Omega_G\Omega_L+\Omega_G^2+\frac{g_{tt}}{g_{\phi\phi}}=0\,,
\end{eqnarray}
in which
\begin{eqnarray}
    \gamma=\frac{1}{g_{\phi\phi}}\left(\frac{g_{\phi\phi,\theta}}{2\beta_EF_{\theta\phi}}\right)^2.
\end{eqnarray}
Finally, solving Eq. (\ref{eq.exp}) and using expression for gravitational frequency $\Omega_G^2$ given in Eq. (\ref{eq.Omega ch}) paves a way to find Larmor frequency $\Omega_L$ and angular velocity $\Omega_E$ for charged particle as:
\begin{subequations}\label{eq.Omega ch.3}
    \begin{align}
        &\Omega_L=\frac{1}{1+\gamma}\left[-\sqrt{-\frac{g_{tt,r}}{g_{\phi\phi,r}}}+\sqrt{\gamma\frac{g_{tt,r}}{g_{\phi\phi,r}}-(1+\gamma)\frac{g_{tt}}{g_{\phi\phi}}}\right]\,,\\
        &\Omega_E=\frac{1}{1+\gamma}\left[\gamma\sqrt{-\frac{g_{tt,r}}{g_{\phi\phi,r}}}+\sqrt{\gamma\frac{g_{tt,r}}{g_{\phi\phi,r}}-(1+\gamma)\frac{g_{tt}}{g_{\phi\phi}}}\right]\,.
    \end{align}
\end{subequations}
In stationary point, which is located at the equatorial plane $\theta={\pi}/{2}$, Eq.~(\ref{eq.Omega ch.3}) takes the form for the test particles in the vicinity of the Schwarzschild-BR BH:
\begin{eqnarray}\label{eq.Omega ch.4}
    \Omega_E^*=\Omega_G^*=\frac{\left(1+B^2r^2\right)}{r}\sqrt{\frac{M}{r}}\,,
\end{eqnarray}
which converts to the form Keplerian frequency of the test particle orbiting Schwarzschild BH $\Omega_K=\frac{1}{r}\sqrt{\frac{M}{r}}$ for the case $B\to0$.

To derive epicyclic frequencies ($\Omega_r, \Omega_\theta$) of the charged particles we have to find the function $V_E(r,\theta)$ using normalization condition $u_\mu u^\mu=-1$:

\begin{eqnarray}\label{eq.four-vel. ch.}
    g_{rr}\Dot{r}^2+g_{\theta\theta}\Dot{\theta}^2+V_E(r,\theta)=0\,,
\end{eqnarray}
in which
\begin{eqnarray}\label{eq.V. ch}
    V_E(r,\theta)=1+\frac{\mathcal{E}^2}{g_{tt}}+\frac{(l-\beta_EA_\phi)^2}{g_{\phi\phi}}\,,
\end{eqnarray}
here we have employed Eq.~(\ref{eq.motion of ch.}). Similarly, what we perform in Sec.~\ref{sec:Epicyclic_Frequencies_for_Bparticles_in_SBR_BH}, considering small perturbation ($r_0,\theta_0$) near a circular orbit $V_E(r,\theta)$ can be expended as:
\begin{eqnarray}\label{eq.exp}
    V(r,\theta)\approx\frac{1}{2}\partial_r^2V(r,\theta)\vert_{r_0,\theta_0}\delta r^2+\frac{1}{2}\partial_\theta^2V(r,\theta)\vert_{r_0,\theta_0}\delta\theta^2,
\end{eqnarray}
here again we take stability conditions into account. Subsequently, Eqs.~(\ref{eq.motion of ch.},\ref{eq.V. ch}, \ref{eq.exp}) yields to the independent harmonic oscillator equations for electrically charged particles:
\begin{eqnarray}
\frac{d^2}{dt^2} \delta r + \Omega_r^2 \delta r = 0, \quad \frac{d^2}{dt^2} \delta \theta + \Omega_\theta^2 \delta \theta = 0,
\end{eqnarray}
where $\Omega_r^2$ and $\Omega_\theta^2$ are
\begin{eqnarray}\label{eq.frequency}
\Omega_r^2 = \frac{1}{2g_{rr}\dot{t}^2} \partial_r^2V_E(r,\theta), \quad \Omega_\theta^2 = \frac{1}{2g_{\theta\theta}\dot{t}^2}  \partial_\theta^2V_E(r,\theta).
\end{eqnarray}

Then Eq.~(\ref{eq.V. ch}) enables to find expressions for electrically charged particles in the vicinity of the Schwarzschild-BR BH:
\begin{widetext}
    \begin{subequations}
        \begin{align}\label{eq.epc.}
            &\Omega_r^2=\frac{1}{g_{rr}}\left[\frac{g_{tt,r}^2}{g_{tt}}-\frac{1}{2}g_{tt,rr}+\Omega_E^2\left(\frac{g^2_{\phi\phi,r}}{g_{\phi\phi}}-\frac{1}{2}g_{\phi\phi,rr}\right)-\frac{\beta_E\Omega_E}{\Dot{t}}\left(A_{\phi,rr}-2A_{\phi,r}\frac{g_{\phi\phi,r}}{g_{\phi\phi}}\right)-\frac{\beta_E^2A^2_{\phi,r}}{\Dot{t}^2g_{\phi\phi}}\right],\\\label{eq.epc.1}
            &\Omega_\theta^2=\frac{1}{g_{\theta\theta}}\left[\frac{g_{tt,\theta}^2}{g_{tt}}-\frac{1}{2}g_{tt,\theta\theta}+\Omega_E^2\left(\frac{g^2_{\phi\phi,\theta}}{g_{\phi\phi}}-\frac{1}{2}g_{\phi\phi,\theta\theta}\right)-\frac{\beta_E\Omega_E}{\Dot{t}}\left(A_{\phi,\theta\theta}-2A_{\phi,\theta}\frac{g_{\phi\phi,\theta}}{g_{\phi\phi}}\right)-\frac{\beta_E^2A^2_{\phi,\theta}}{\Dot{t}^2g_{\phi\phi}}\right].
        \end{align}
    \end{subequations}
\end{widetext}
Eqs.~(\ref{eq.epc.},\ref{eq.epc.1}) can be recalculated for electrically charged particle orbiting Schwarzschild-BR BH at the equatorial plane as:
    \begin{subequations}\label{eq.Omega ch.6}
        \begin{align}
            &\Omega_r^2=\Omega^2_{Schw\,r}-\Xi_r(r)+\Theta_r(r)+\Phi_r(r)+\Lambda_r(r)+\mathcal{K}_r(r),\\
            &\Omega_\theta^2=\Omega_{Schw\,\theta}^2+\Xi_\theta(r)-\Theta_\theta(r)+\Phi_\theta(r)\,,
        \end{align}
    \end{subequations}
where new variables are given in Appendix~\ref{sec.app} and $\Omega_{Schw\,r}^2=\frac{M (r-6 M)}{r^4}$, $\Omega_{Schw\,\theta}=\frac{M}{r^3}$ are the epicyclic frequencies of the tests particles in the vicinity of the Schwarzschild BH.

\begin{figure*}[ht!]\centering
\includegraphics[width=0.45\textwidth]{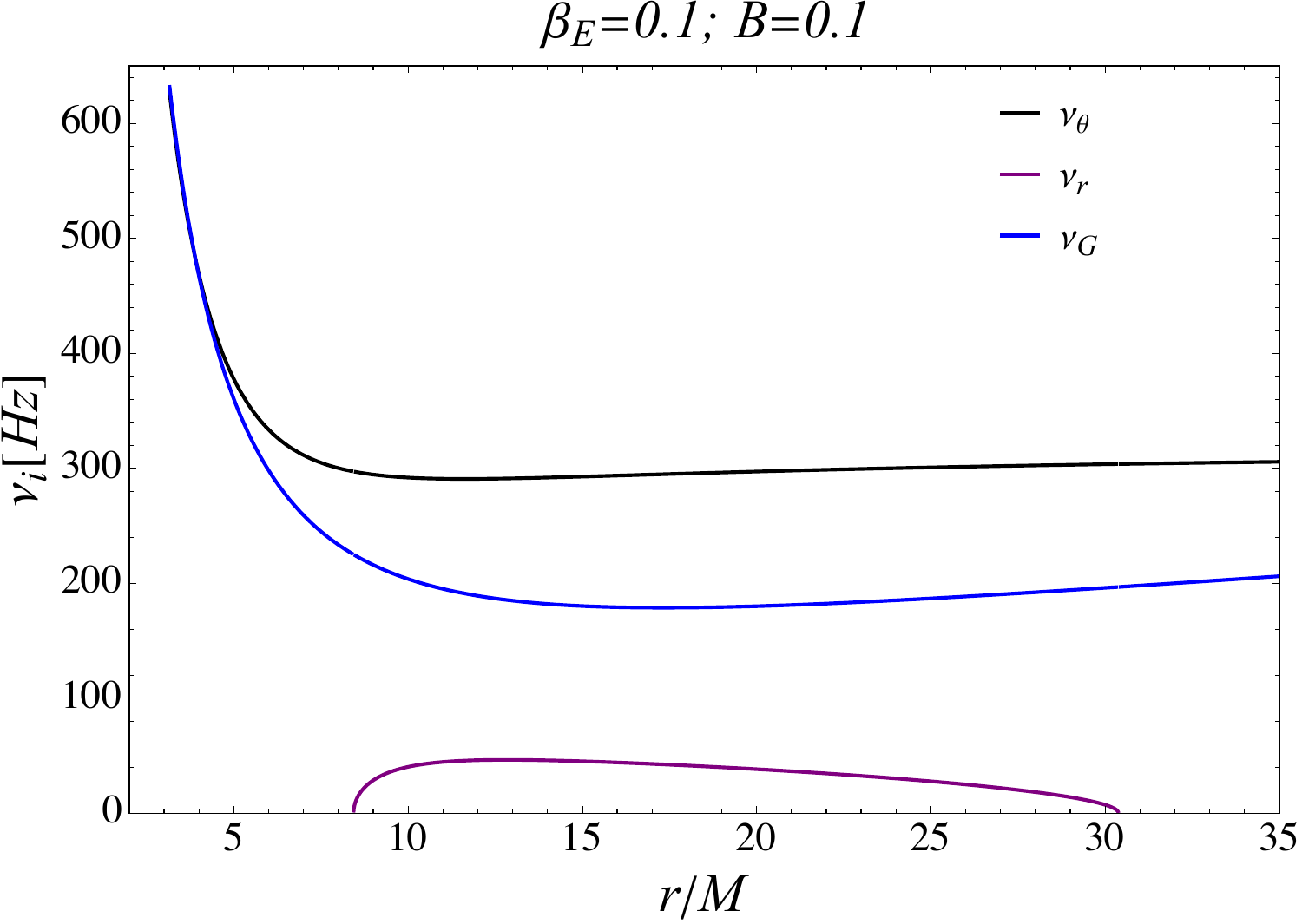}
\includegraphics[width=0.45\textwidth]{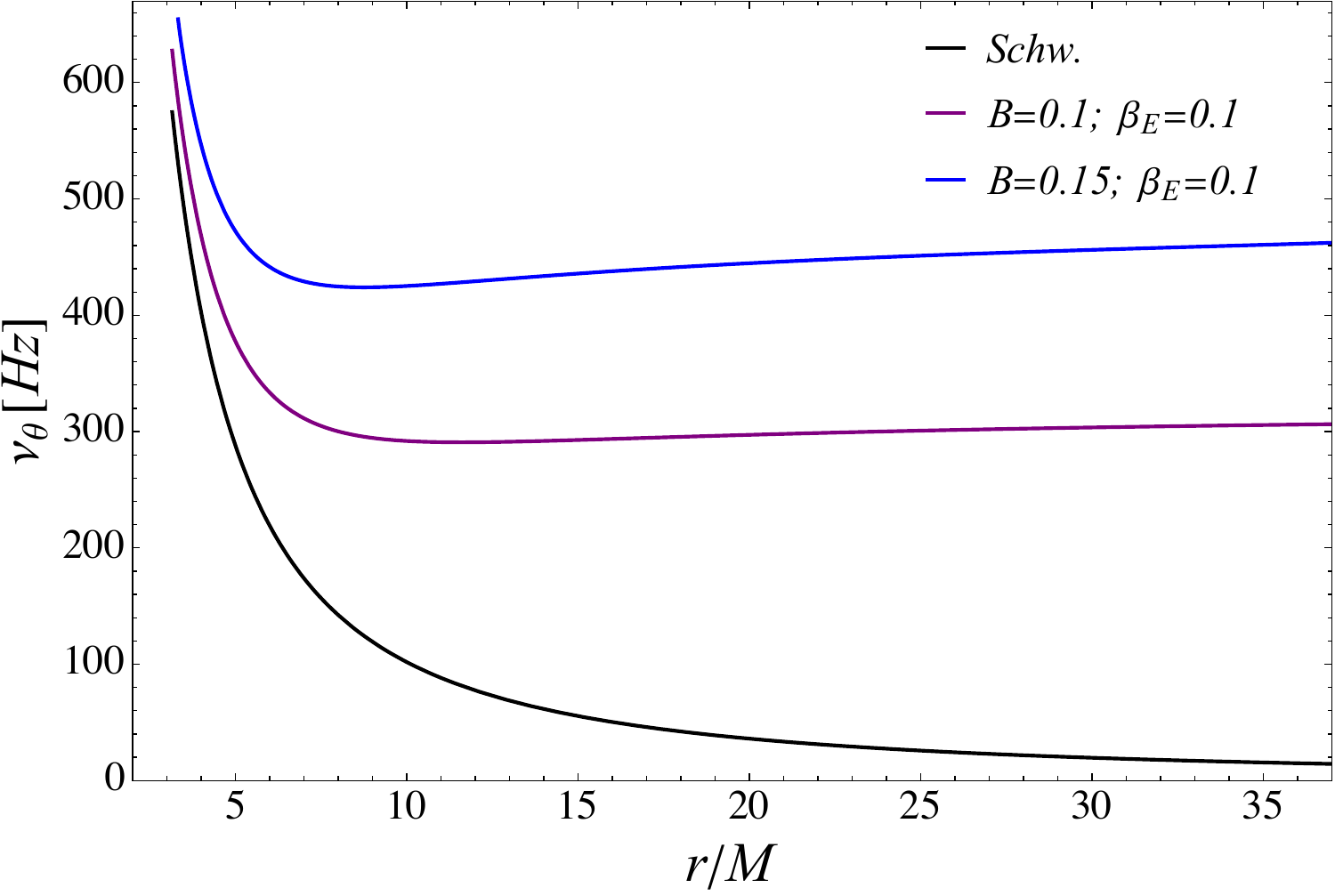}
\includegraphics[width=0.45\textwidth]{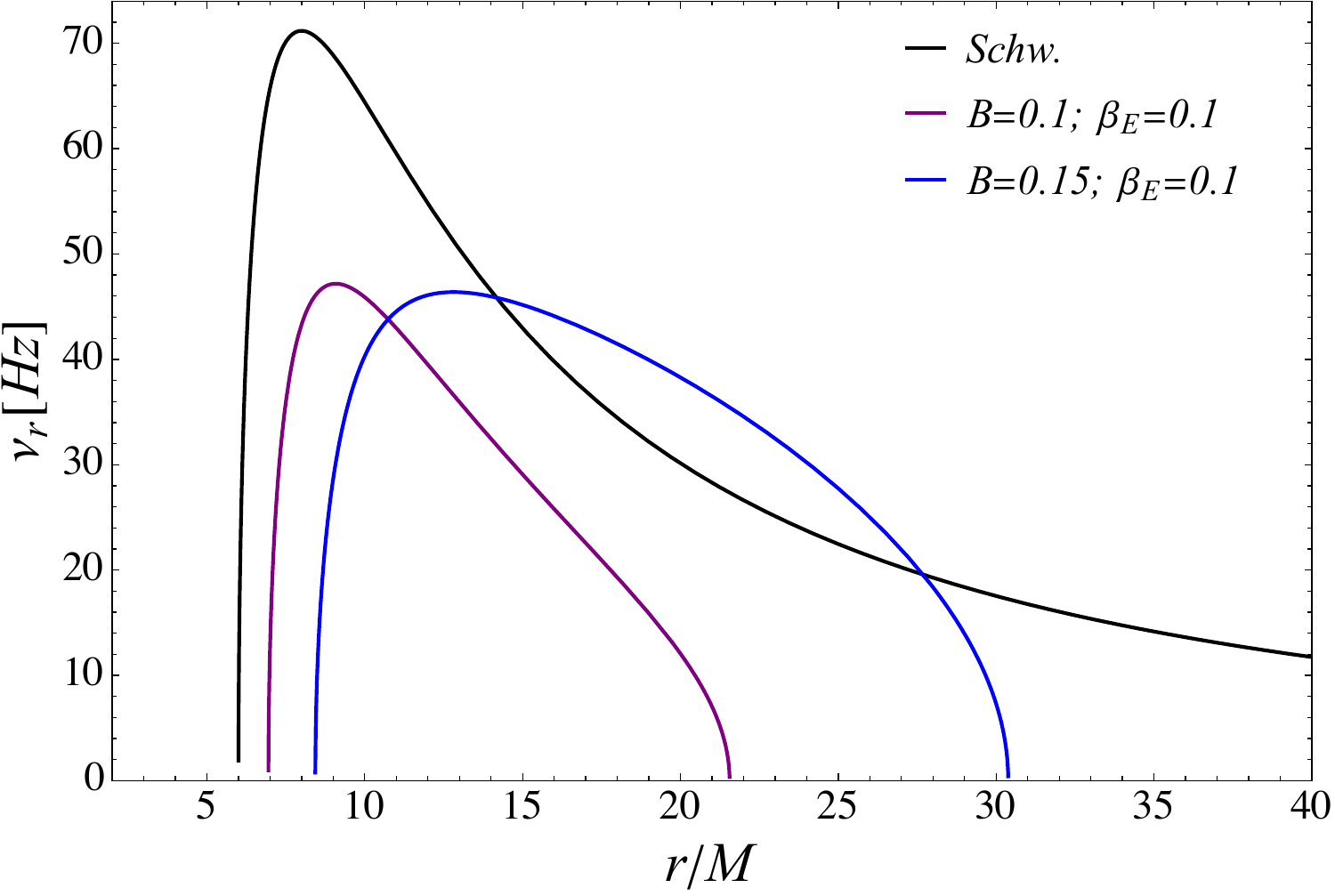}
\includegraphics[width=0.45\textwidth]{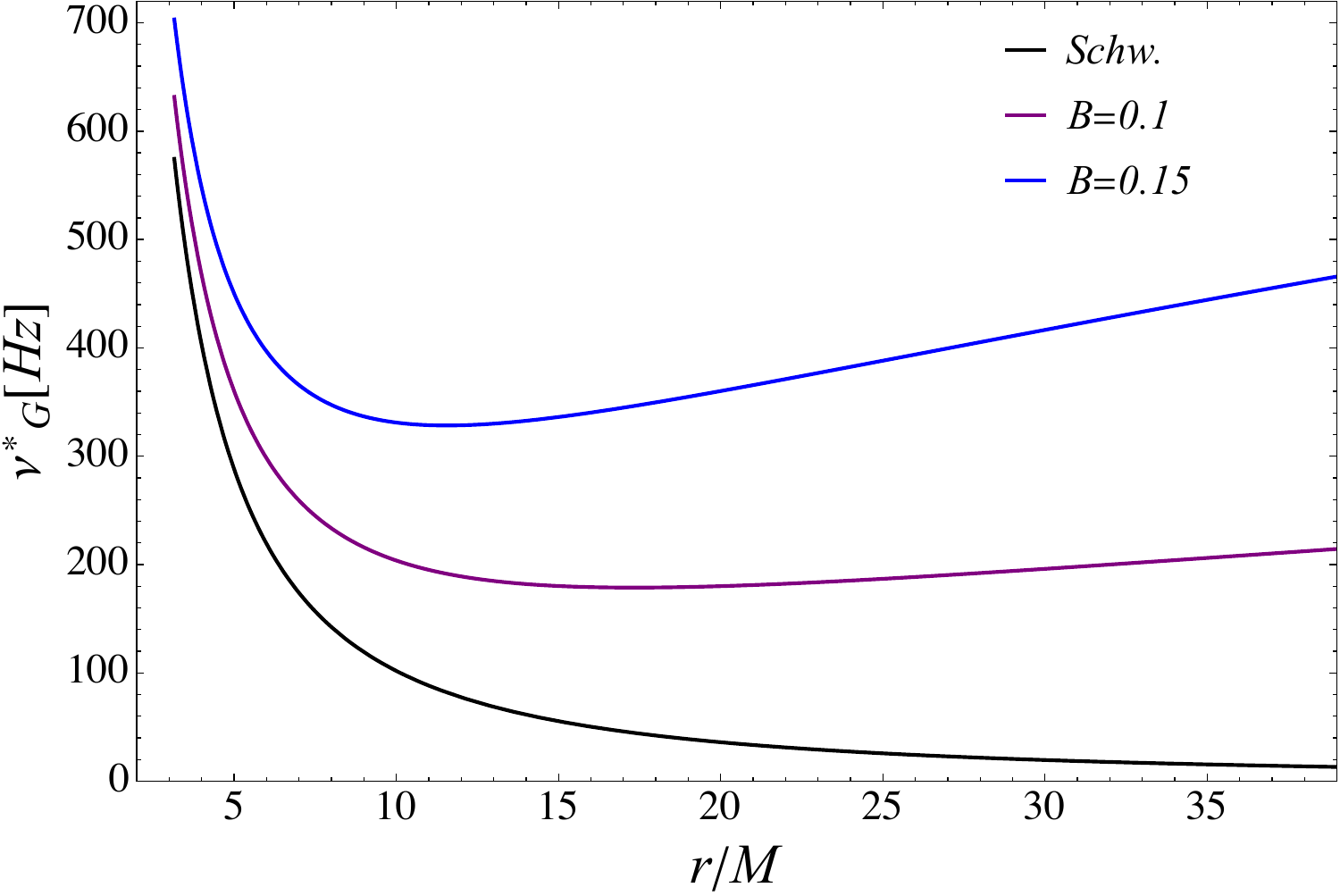}
\caption{Radial dependence of the small harmonic oscillation frequencies ($\nu_r,\,\nu_\theta,\,\nu_G^*$) of the electrically charged particles in the vicinity of the Schwarzschild-BR BH for different cases. Here we should note that Eq.(\ref{eq.nu}) is employed to recover physical units.\label{Fig.frequency ch.}}
\end{figure*}

Then we illustrate how the frequencies ($\nu_r,\,\nu_\theta,\,\nu_G^*$) of the electrically charged particles depend on the parameters $B$, $\beta_E$ in Fig.~\ref{Fig.frequency ch.}. One can notice from plots in Fig.~\ref{Fig.frequency ch.} that increasing the value of the magnetic field $B$ causes growing the values of the $\nu_\theta$, $\nu_G^*$, however the peaks of the radial frequency $\nu_r$ decreasing with the growing the value of the $B$.

\subsection{Charged particle trajectory dynamics interpretation}\label{sec:charged_particle_trajectory_dynamics_interpretation}

We study the trajectory dynamics, of charged particles, and we provide the profiles of their parametrisation as follows.

\begin{figure*}[ht!]
\includegraphics[width=0.8\textwidth]{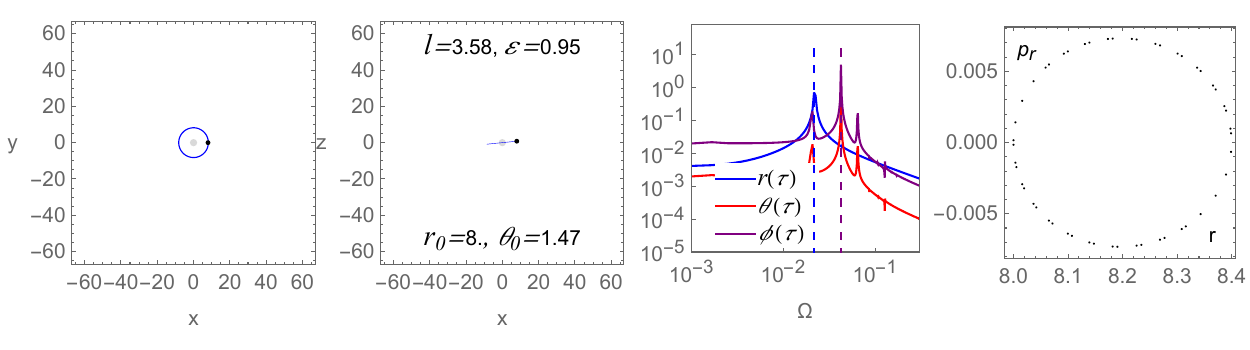}
\includegraphics[width=0.8\textwidth]{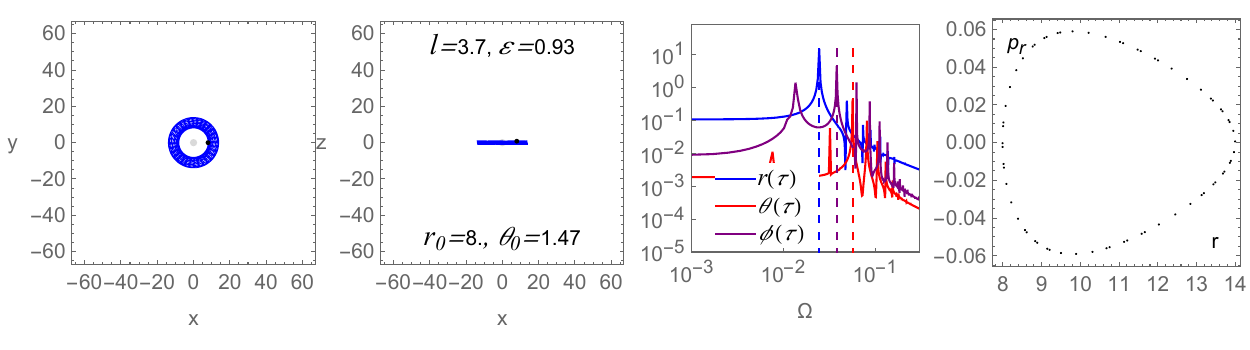}
\includegraphics[width=0.8\textwidth]{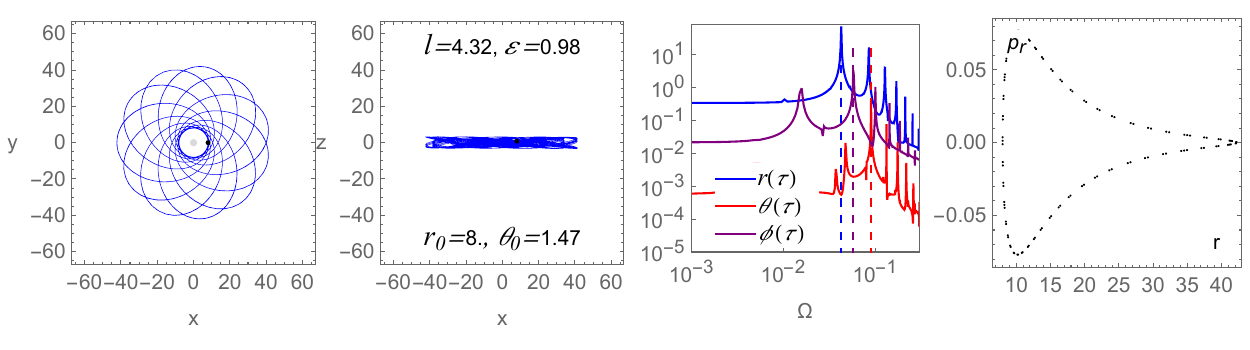}
\caption{The trajectory of the charged particles (first and second column). Power Spectral Density for three $r(\tau)$ (blue), $\theta(\tau)$ (red), $\phi(\tau)$ (purple) coordinates along the particle’s trajectory (third column) and corresponding PS (last column) for different values of the magnetic field $B$ (from above row: $B=0, 0.06, 0.11$),here we take the value of the charge as $\beta_E = 0.6$. [See Section
\ref{sec:charged_particle_trajectory_dynamics_interpretation} for details]
\label{Fig.trajectory ch.1}}
\end{figure*}

\begin{figure*}[ht!]
\includegraphics[width=0.8\textwidth]{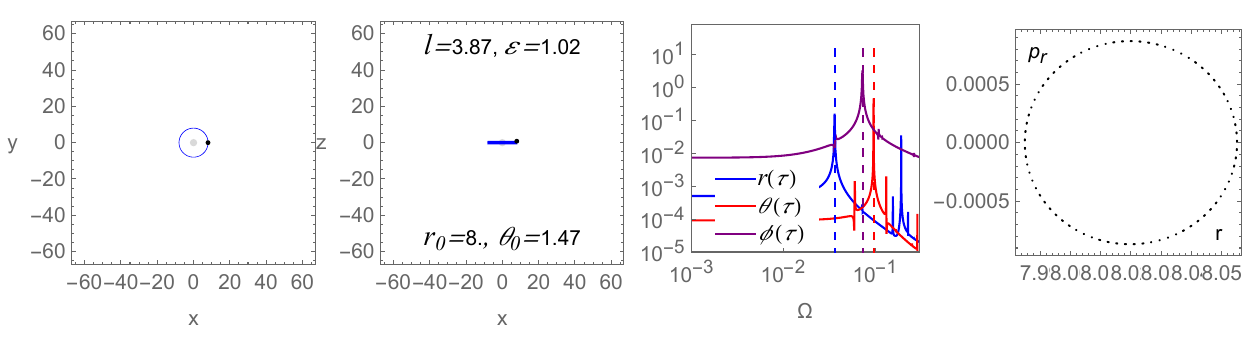}
\includegraphics[width=0.8\textwidth]{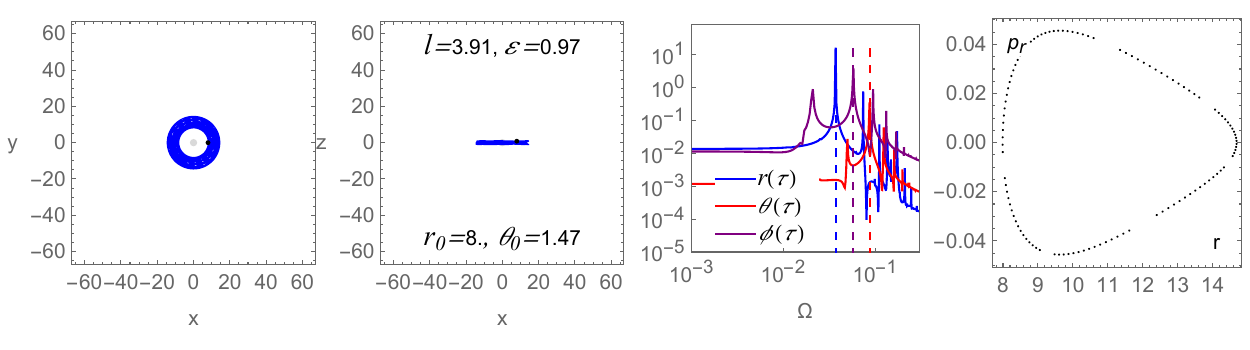}
\includegraphics[width=0.8\textwidth]{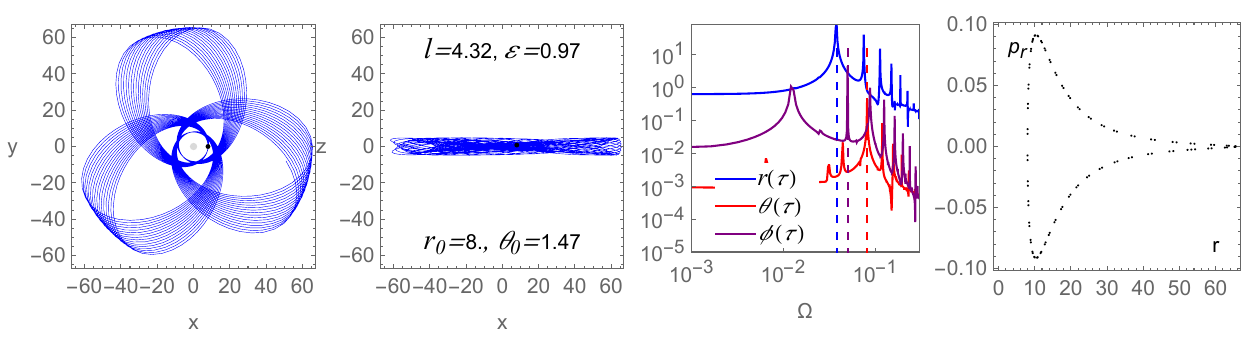}
\caption{The trajectory of the charged particles (first and second column). Power Spectral Density for three $r(\tau)$ (blue), $\theta(\tau)$ (red), $\phi(\tau)$ (purple) coordinates along the particle’s trajectory (third column) and corresponding PS (last column) for different values of the charge of the particle $q$ (from above row: $q=0, 0.4, 0.7; B=0.1$),here we take the value of the magnetic field strength as $B = 0.1$.[See Section
\ref{sec:charged_particle_trajectory_dynamics_interpretation} for details]
\label{Fig.trajectory ch.2}}
\end{figure*}

Figures~\ref{Fig.trajectory ch.1} and \ref{Fig.trajectory ch.2} have the same basic characteristics as Fig.~\ref{fig:Trajectory_for_magnetized}, i.e. they present trajectory data for magnetized particles in a \(3 \times 4\) panel matrix. Each row corresponds to a set of 4 panel, and each row corresponds to a different magnetic field, $B$ change. 

In Fig.~\ref{Fig.trajectory ch.1}, each row corresponds to a panel with a different  magnetic field strength \(B=0\) (first row), \(B=0.06\) (second row), and \(B=0.11\) (third row). This case corresponds to the initial radial coordinate \(r_0=8.0\), and initial angle \(\theta_0=1.47\).

However, unlike in Fig.~\ref{fig:Trajectory_for_magnetized}, Fig.~\ref{Fig.trajectory ch.1} presents the trajectory of a charged particle, with $\beta_E=0.6$.

As the magnetic field increases, both \(l\) and \(\epsilon\) increase accordingly: \((l, \epsilon) = (3.58, 0.95)\) for \(B=0\), \((3.7, 0.93)\) for \(B=0.06\), and \((4.32, 0.98)\) for \(B=0.11\). The small decrease that we see in the energy as the magnetic field increases, it is because the charge and the magnetic field, have the correct combination inside the effective potential, which makes the energy a bit lower.

For each column, of Fig.~\ref{Fig.trajectory ch.1}, we observe the following.

\paragraph{Column 1:}
In all cases, the particle follows a circular, epicyclic trajectory. The paths become more densely packed as the magnetic field increases, as expected, and as observed in the case of the non-charged particle trajectories, shown in Fig.~\ref{fig:Trajectory_for_magnetized}, in Section (\ref{sec:trajectory_dynamics_interpretation}).

\paragraph{Column 2:}
In the first row (\(B=0\)), the trajectory shows a slight tilt in the \(xz\)-plane. As observed in the case of the non-charged particle trajectories, shown in Fig.~\ref{fig:Trajectory_for_magnetized}, in Section (\ref{sec:trajectory_dynamics_interpretation}), this is due to the initial condition \(\theta_0 = \frac{\pi}{2} + \delta\theta_0\), i.e., a small deviation from the equatorial plane. In the second and third rows, the trajectory stabilizes horizontally, indicating that the magnetic field confines the particle motion near the equatorial plane, suppressing vertical excursions.

\paragraph{Column 3:}
The PSD for the spherical coordinates \(r(\tau)\), \(\theta(\tau)\), and \(\phi(\tau)\) is presented. As observed in the case of the non-charged particle trajectories, shown in Fig.~\ref{fig:Trajectory_for_magnetized}, in Section (\ref{sec:trajectory_dynamics_interpretation}), as the magnetic field increases (from top to bottom row), the main frequency peak shifts to higher values, indicating the increasing influence of the Lorentz force on the characteristic timescales of motion. For regular trajectories oscillating around effective potential minima (e.g., the \(B=0\) case), the PSD shows distinct main peaks corresponding to the fundamental frequencies calculated analytically, along with their higher harmonics (overtones). As the magnetic field increases to intermediate values (\(B=0.06, 0.11\)), the high-frequency region transitions from smooth to increasingly noisy, reflecting the onset of chaotic dynamics. We also observe that the azimuthal motion \(\phi(\tau)\) exhibits the strongest high-frequency components, consistent with the higher curvature of this degree of freedom.

\paragraph{Column 4:}
The radial PS shows chaotic behavior in the absence of a magnetic field. As the magnetic field increases, the phase space becomes more regular, indicating a stabilizing effect due to the magnetic field.

Unlike in Fig.~\ref{Fig.trajectory ch.1}, Fig.~\ref{Fig.trajectory ch.2} presents the trajectory of a charged particle with a fixed magnetic field, and the charge of the particle, is increasing. Each row corresponds to a panel with a different particle charge field strength \(\beta_E=0\) (first row), \(\beta_E=0.4\) (second row), and \(\beta_E=0.7\) (third row), while the magnetic field is fixed to $B=0.1$. This case corresponds to the initial radial coordinate \(r_0=8.0\), and initial angle \(\theta_0=1.47\), as in Fig.~\ref{Fig.trajectory ch.1}.

As the particle charge increases, \(l\) increases and \(\epsilon\) decreases. In particular, \((l, \epsilon) = (3.87, 1.02)\) for \(\beta_E=0\), \((3.91, 0.97)\) for \(\beta_E=0.4\), and \((4.32, 0.97)\) for \(\beta_E=0.7\).

For each column, of Fig.~\ref{Fig.trajectory ch.2}, we observe the following.

\paragraph{Column 1:}
In all cases, the particle follows a circular, epicyclic trajectory. The paths become more densely packed as the magnetic field increases, as expected, and as observed in the case of the non-charged particle trajectories, shown in Fig.~\ref{fig:Trajectory_for_magnetized}, in Section (\ref{sec:trajectory_dynamics_interpretation}). We also observe a special kind of taxonomy in the last plane which is a byproduct of the specific energy and angular momentum of the system as was demonstrated by \citet{Uktamov:2024zmj}.

\paragraph{Column 2:}
In the first row (\(\beta_E=0\)), the trajectory shows a slight tilt in the \(xz\)-plane. As observed in the case of the non-charged particle trajectories, shown in Fig.~\ref{fig:Trajectory_for_magnetized}, in Section (\ref{sec:trajectory_dynamics_interpretation}), this is due to the initial condition \(\theta_0 = \frac{\pi}{2} + \delta\theta_0\), i.e., a small deviation from the equatorial plane. In the second and third rows, the trajectory stabilizes horizontally, indicating that the magnetic field confines the particle motion near the equatorial plane, suppressing vertical excursions, as done in the previous cases.

\paragraph{Column 3:}
The PSD for the spherical coordinates \(r(\tau)\), \(\theta(\tau)\), and \(\phi(\tau)\) is presented. As observed in the case of the magnetized particle trajectories, shown in Fig.~\ref{fig:Trajectory_for_magnetized}, in Section (\ref{sec:trajectory_dynamics_interpretation}), here we observe that as the particle charge  increases (from top to bottom row), the main frequency peak shifts to higher values, indicating the increasing influence of the Lorentz force on the characteristic timescales of motion. For regular trajectories oscillating around effective potential minima (e.g., the \(\beta_E=0\) case), the PSD shows distinct main peaks corresponding to the fundamental frequencies calculated analytically, along with their higher harmonics (overtones). As the particle charge increases to intermediate values (\(\beta_E=0.4, 0.7\)), the high-frequency region transitions from smooth to increasingly noisy, reflecting the onset of chaotic dynamics. We also observe that the azimuthal motion \(\phi(\tau)\) exhibits the strongest high-frequency components, consistent with the higher curvature of this degree of freedom.

\paragraph{Column 4:}
The radial PS shows chaotic behavior in the absence of a particle charge. As the particle charge increases, the phase space becomes more regular, indicating a stabilizing effect due to the increase of the particle charged.

\section{Conclusion}\label{sec:conclusion}

In this paper, we start with analyzing the nature of the Schwarzschild-BR BH and show that in the weak field limit, the space time metric (\ref{eq.metric}) exhibits to the Wald's solution~\cite{Wald:1984rg} for the Schwarzschild BH immersed in an external uniform magnetic field.  Then, we have derived the expression for the effective potential $V_{eff}$ of the magnetized particle in (\ref{eq.Veff1})
and the expression for the effective of the electrically charged particle in (\ref{eq.V. ch}). Subsequently, using effective potential Eqs.(\ref{eq.Veff1}, \ref{eq.V. ch}) we have analyzed ISCO parameters $r_{ISCO}$, $\mathcal{E}_{ISCO}$, $l_{ISCO}$ of the magnetized particles and charged particles. 

Then, we have derived explicit expression for the orbital, radial and vertical frequencies of the magnetized particles (\ref{eq.Omega r m},\ref{Omega theta m}) and electrically charged particles (\ref{eq.Omega ch.6}) orbiting around Schwarzschild BR BH. Furthermore, we have depicted radial dependence of the small harmonic oscillation frequencies of the particles with magnetic dipole momentum in Fig.~\ref{Fig.frequency magnetized} and small harmonic oscillation frequencies of the electrically charged particles in Fig.~\ref{Fig.frequency ch.}.

The analysis of particle trajectories reveals several important dynamical features. For magnetized particles, increasing the magnetic field strength $B$ leads to larger specific energy $\mathcal{E}$ and angular momentum $l$, while simultaneously confining the motion closer to the equatorial plane and suppressing vertical excursions. The PSD analysis shows that the characteristic frequencies shift to higher values with increasing $B$, indicating the growing influence of the Lorentz force. Notably, the radial phase space transitions from chaotic behavior in the absence of a magnetic field to increasingly regular motion as $B$ increases, demonstrating a stabilizing effect of the magnetic field on particle dynamics. Similar behavior is observed for charged particles: increasing either the magnetic field $B$ or the particle charge $q$ enhances the confinement near the equatorial plane and regularizes the trajectories. However, while $l$ increases monotonically with both $B$ and $q$, the specific energy $\mathcal{E}$ exhibits a non-monotonic dependence on the magnetic field for charged particles, decreasing slightly at intermediate field strengths before increasing. The PSD analysis confirms that both stronger magnetic fields and higher particle charges shift the frequency peaks upward and promote more ordered motion, with the azimuthal degree of freedom consistently showing the strongest high-frequency components due to its higher curvature. These results collectively establish that both the external magnetic field and the particle's electromagnetic properties play crucial roles in determining the stability and regularity of orbits around Schwarzschild-BR BH.


\section*{ACKNOWLEDGEMENT}

This research was funded by the National Natural Science Foundation of China (NSFC) under Grant No. U2541210.
\appendix
\section{Useful expressions}\label{sec.app}

We present useful expression below.

The variables for Eqs.(\ref{eq.dr},\ref{eq.l E}) are :
\begin{widetext}
\begin{align}
    &\mathcal{V}_1=f(r)^2
        \left[\frac{l^2N}{r^2}+(1+\beta f(r))^2\right]\,,\\
    &\mathcal{V}_2=\left(1-\frac{2M}{r}-B^2M^2\right)\left(1+\frac{(1+B^2r^2)(l+\frac{\beta_E(1-\frac{1}{\sqrt{1+B^2r^2}})^2}{B})^2}{r^2}\right)\,,\\
    &\mathcal{Y}=r-3M-B^2 M^2 r -M B^2 r^2\,,\\
    &\mathcal{Z}=\bigg(-M +r\bigg)\bigg(1 +B^2 M r\bigg)\,.
\end{align}
\begin{align}
    &\mathcal{A}=\sqrt{B^2 \beta_E^2\left(4M^2+B^2 M^3 r+r^2-4Mr-MB^2 r^3\right)+\frac{4M}{r^2}\mathcal{Y}}\,,\\
    &\mathcal{B}_1=4M^2+4B^2 M^3 r-4Mr+(1-B^2 M^2)^2 r^2+B^4 M^3 r^3\,,\\
    &\mathcal{B}_2=2B\beta_E r^2\left(4Mr-B^2 M^3 r-r^2-4M^2+M^2 B^2 r^2\right)\,,\\
    &\mathcal{B}_3=B^2\beta_E^2 r^2\left(3B^4 M^ 5r^2-2r^3-B^2 r^5+15B^2 M^4 r+5B^4 M^4 r^3+16M^3-20M^2 B^2 r^3\right)\,,\\
    &\mathcal{B}_4=B^2\beta_E^2 r^2\left(5M^3 B^2 r^2-2M^3 B^4 r^4+12Mr^2+9MB^2 r^4+MB^4 r^6-24M^2 r-3B^4 M^2 r^5\right)\,.
\end{align}
\end{widetext}
The variables for Eq.(\ref{eq.Omega ch.6}) are
\begin{widetext}
\begin{subequations}
    \begin{align}               &\Xi(r)=B^6M^3\left[r+\beta_E(M+r)\right]\,,\\
            &\Theta_r(r)=\frac{B^4 M \left[-\left(\left[5\beta_E ^2+2\right] M^2\right)-6 M r+\left(\beta_E^2+1\right) r^2\right]}{r}\,,\\
            &\Phi_r(r)=-\frac{B^2 \left[\beta_E^2 r (r-3 M) (r-2 M)+M \left(M^2+12 M r-2 r^2\right)\right]}{r^3}\,,\\
            &\Lambda_r(r)=\frac{\beta  B \left(6 B^2 M r^2-2 B^2 r^3+6 M-3 r\right) \Omega_G^* \sqrt{B^2 (-M) (M+r)-\frac{3 M}{r}+1}}{r \sqrt{B^2 r^2+1}}\,,\\
            &\mathcal{K}_r(r)=\frac{\beta_EB^3 M \left(M \left(2 B^2 r^2+3\right)-2 r\right) \sqrt{\frac{M \left(B^2 r^2+1\right)^2}{r^3}} \sqrt{B^2 (-M) (M+r)-\frac{3 M}{r}+1}}{\sqrt{B^2 r^2+1}}\,,\\
            &\Xi_\theta(r)=B^6 M^4+\frac{3 B^4 M^3}{r}+\frac{B^2 \left(-M^3+6 M^2 r-3 M r^2+r^3\right)}{r^3}\,,\\
            &\Theta_\theta(r)=\frac{\beta_EB \Omega_G^* \sqrt{-B^2 M(M+r)-\frac{3 M}{r}+1}}{\sqrt{B^2 r^2+1}}\,,\\
            &\Phi_\theta(r)=\frac{\beta_EB^3 M (M-2 r) \Omega_G^* \sqrt{B^2 (-M) (M+r)-\frac{3 M}{r}+1}}{\sqrt{B^2 r^2+1}}\,,
  \end{align}
\end{subequations}
\end{widetext}
\allowdisplaybreaks

\bibliography{prd/main}    
\end{document}